\RequirePackage{fix-cm}
\documentclass[smallextended]{svjour3}       
\smartqed  
\usepackage{graphicx}
\usepackage{balance}
\usepackage{amsmath,amssymb,amsfonts}
\usepackage{algorithmic}
\usepackage{graphicx}
\usepackage{textcomp}
\usepackage{xcolor}
\def\BibTeX{{\rm B\kern-.05em{\sc i\kern-.025em b}\kern-.08em
    T\kern-.1667em\lower.7ex\hbox{E}\kern-.125emX}}

\usepackage{multirow}
\usepackage{listings}
\usepackage[english=american]{csquotes}
\usepackage{float}
\usepackage{soul}
\usepackage{adjustbox}
\usepackage{tcolorbox}
\usepackage{rotating}

\usepackage{caption}
\usepackage{subcaption}
\usepackage[hyphens]{url}
\usepackage{hyperref}
\usepackage[numbers, sort]{natbib}

\newcommand{\eg}{\textit{e}.\textit{g}., }
\newcommand{\ie}{\textit{i}.\textit{e}., }

\definecolor{asparagus}{rgb}{0.53, 0.66, 0.42}

\begin{document}

\title{Silent Bugs in Deep Learning Frameworks: An Empirical Study of Keras and TensorFlow\thanks{This work was supported by: Fonds de Recherche du Québec (FRQ), the Canadian Institute for Advanced Research (CIFAR) as well as the DEEL project CRDPJ 537462-18 funded by the National Science and Engineering Research Council of Canada (NSERC) and the Consortium for Research and Innovation in Aerospace in Québec (CRIAQ), together with its industrial partners Thales Canada inc, Bell Textron Canada Limited, CAE inc and Bombardier inc.}
}

\titlerunning{Silent Bugs in Deep Learning Frameworks}        

\author{Florian Tambon*         \and
        Amin Nikanjam* \and
        Le An \and
        Foutse Khomh \and
        Giuliano Antoniol
}

\authorrunning{F. Tambon et al.} 

\institute{Florian Tambon, Amin Nikanjam, Le An, Foutse Khomh, Giuliano Antoniol \at
              Polytechnique Montreal, Montreal, QC H3C 3A7, Canada\\
              \email{\{florian-2.tambon, amin.nikanjam, le.an, foutse.khomh, giuliano.antoniol\}@polymtl.ca}\\
              "*" authors contributed equally
}

\date{Received: date / Accepted: date}

\maketitle

\begin{abstract}
Deep Learning (DL) frameworks are now widely used, simplifying the creation of complex models as well as their integration into various applications even among non-DL experts. However, like any other programs, they are prone to bugs. This paper deals with the subcategory of bugs named \emph{silent bugs}: they lead to wrong behavior but they do not cause system crashes or hangs, nor show an error message to the user. Such bugs are even more dangerous in DL applications and frameworks due to the \enquote{black-box} and stochastic nature of the DL systems (\ie the end user can not understand how the model makes decisions). This paper presents the first empirical study of the \emph{silent bugs} in Tensorflow, specifically its high-level API Keras, and their impact on users' programs. We extracted closed issues related to Keras API from the TensorFlow GitHub repository. Out of the 1,168 issues that we gathered, 77 were reproducible silent bugs affecting users' programs. We categorized the bugs based on the effects on the users' programs and the components where the issues occurred, using information from the issue reports. We then derived a threat level for each of the issues, based on the impact they had on the users' programs. To assess the relevance of identified categories and the impact scale, we conducted an online survey with 103 DL developers. The participants generally agreed with the significant impact of silent bugs in DL frameworks and how they impact users and acknowledged our findings (i.e., categories of silent bugs and the proposed impact scale).
\end{abstract}

\section{Introduction} 

Deep Learning (DL) frameworks such as TensorFlow \cite{TF,abadi2016tensorflow}, Keras \cite{keras}, Pytorch \cite{torch}, MLlib \cite{MLlib-Spark}, and Jax \cite{jax} are becoming more and more popular to develop sophisticated and complex applications in various domains; from face recognition, fraud detection in financial companies, diagnosis and treatment of diseases, and natural language processing to autonomous vehicles. While users rely on these frameworks and develop their DL applications on top of them, they are subject to {\em bugs} like any other software. Bugs in DL frameworks are essential because a bug in a framework can lead to bugs in any application that leverages that framework \cite{jia2021symptoms}. Many of these bugs never make it into released versions, others manifest themselves in easy-to-detect ways (\eg{} error messages, program crashes, or never-ending hangs) and some non-breaking ones remain silent leading to wrong calculations in DL models, training, or inference without any obvious symptoms for the user. We refer to  \textit{silent bugs} as bugs that result in wrong computation or behavior of the model compared to its expected outcome, without apparent symptoms like crash, hang, or an error message. 

We surmise \textit{silent bugs} are especially insidious in DL frameworks. Indeed, finding such bugs in DL frameworks can be very hard and tedious as unit tests are limited at finding such defects \cite{Jia21-icsme} and they can lead to severe impact on users' DL applications built on top of such frameworks. Even with bugs in the DL frameworks, the model of the DL applications may still be created, trained, and appeared to work. A decrease in computation time, prediction accuracy or even a user interface with incorrect information is possible but it is difficult for a user to link a suspicious result/behavior to a concrete bug in DL frameworks, \eg{} wrong update of weights in neural networks during the training. More generally, because of the lack of concrete error, the users may assume the problem lies in their code. 
Hence, any bugs in DL frameworks may lead to bugs in the user's program that the user might not, easily, become aware of (because of the lack of observable error). And even if they become aware of the issue, they might not directly suspect the error is within the DL framework itself and not their own code.

As a motivating example, Figure \ref{fig:motivatingExample} illustrates a bug reported on TensorFlow repository over GitHub. When saving a trained model and loading it again, the accuracy drops to its default value. After training, the model showed 100\% accuracy and the loss dropped to zero. Then, the user saved the model to the disk and immediately loaded it back, and tested the accuracy again, which dropped to 50\%. As the expected behavior, the accuracy should remain the same after loading a previously trained and saved model. The bug was acknowledged by reproducing the issue and providing a gist over Google Colab (a gist is 
\enquote{a simple way to share snippets and pastes with others} \cite{wang2015gist}). In the user-provided example, the model and data are very simple and only used to demonstrate the impact of the reported bug. This bug is related to weights modification when reloading the model, which has been fixed later and the issue was closed. We have classified this bug as "Wrong save/reload" in our study.

\begin{figure}
\centering
\begin{adjustbox}{left,width=0.95\columnwidth}
\begin{lstlisting}[frame = single]
import tensorflow as tf

model = tf.keras.models.Sequential()
model.add(tf.keras.layers.Dense(units=2, activation='softmax', name='output'))
model.compile(optimizer=tf.keras.optimizers.Adam(lr=10),
              loss='sparse_categorical_crossentropy',
              metrics=['accuracy'])
dummy_data_x = [[0, 0], [1, 0], [0, 1], [1, 1]]
dummy_data_y = [0, 1, 0, 1]
print(model.evaluate(x=dummy_data_x, y=dummy_data_y))
model.fit(x=dummy_data_x, y=dummy_data_y, epochs=10)
print(model.evaluate(x=dummy_data_x, y=dummy_data_y))
model.save('test_model')
model = tf.keras.models.load_model('test_model')
print(model.evaluate(x=dummy_data_x, y=dummy_data_y))
\end{lstlisting}
\end{adjustbox}

\begin{adjustbox}{left,width=0.95\columnwidth}
\begin{lstlisting}
Output:
Before training:
1/1 [===============] - 0s 0s/step - loss: 0.9013 - accuracy: 0.5000
[0.9013183116912842, 0.5]
After training:
1/1 [===============] - 0s 0s/step - loss: 0.0000e+00 - accuracy: 1.0000
[0.0, 1.0]
After loading:
1/1 [===============] - 0s 1000us/step - loss: 0.0000e+00 - accuracy: 0.5000
[0.0, 0.5]
\end{lstlisting}
\end{adjustbox}
\caption{A reported issue over GitHub \cite{issue1}. The bug is silent and has a significant impact on the accuracy of the learned model, classified as "Wrong save/reload" in our study. }
\label{fig:motivatingExample}
\end{figure}

There are several empirical studies on bugs in DL software developed using various frameworks \cite{DL_bugs_1,DL_bugs_2,DL_faults}. Researchers also investigated symptoms, root causes, and repair patterns of bugs inside TensorFlow \cite{jia2021symptoms,jia2020empirical} and their fault-triggering conditions \cite{du2020fault}. Although not directly studying silent bugs, some of the bugs they collected are classified into categories such as \textit{Functional Error} or \textit{Performance Degradation} which could be potentially counted as silent bugs. Nonetheless, they do not precisely study those bug categories and how they impact the users’ code as we present in this study. Other studies highlighted the effect of similar bugs in DL: either that some injected bugs would not necessarily output observable error and even not alter the accuracy of DL models \cite{Jia22-san} or that unit tests of DL framework would not catch as easily mutations affecting parameters value \cite{Jia21-icsme}. Those studies do not focus on existing bugs inside DL framework and instead rely on mutations to probe either DL applications or unit tests of DL framework. Though they are interesting substitutes for real faults, mutations do not necessarily represent exactly actual faults in a system \cite{Papadakis18} which can hamper practical analysis and recommendations for users and developers of such frameworks. Moreover, we also focused, contrary to previous studies, on the impact of said \textit{silent bugs} over users’ code by validating our empirical analysis with a survey. As such, we are not aware of existing studies that specifically targeted real \textit{silent bugs}, bugs that do not result in obvious symptoms or unexpected behaviors, reported by users inside DL frameworks and how they affect said users. Therefore, we formulate our research questions as follows:
 
\begin{itemize}
\item [\textbf{RQ1}] What are the impacts of \textit{silent bugs} inside DL frameworks on DL programs developed by users? What are the common consequences of \textit{silent bugs}?
\item [\textbf{RQ2}] From which part of the framework's API do the \textit{silent bugs} come? What are their main root causes?
\item [\textbf{RQ3}] To what extent are the considered \textit{silent bugs} relevant for DL users? 
\end{itemize}

\begin{figure}[t]
\centering
    \includegraphics[width=0.95\columnwidth]{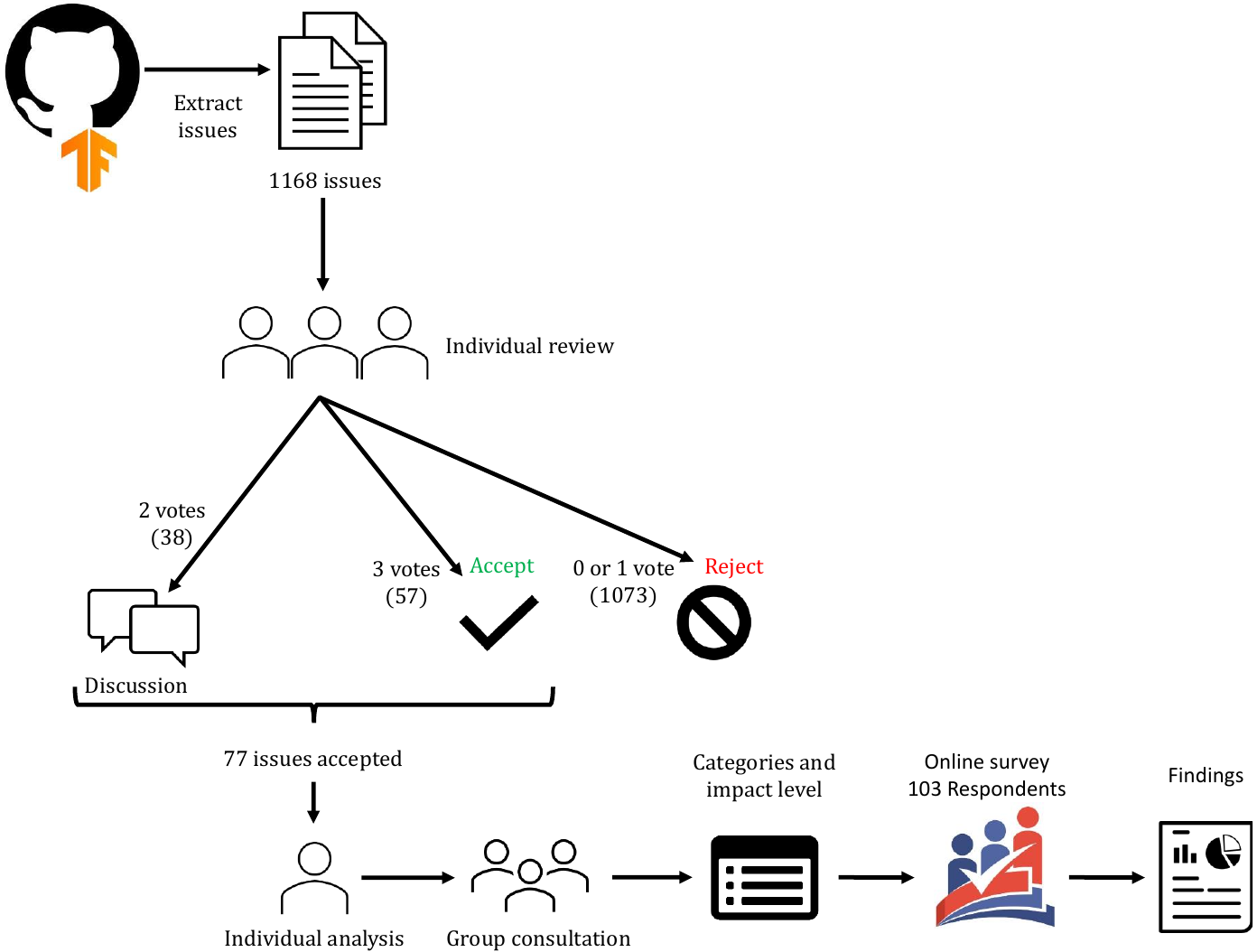}
    \caption{The main steps of our study.}
    \label{fig:process}
    \vspace{-2em}
\end{figure}

In this paper, we conducted an empirical study specifically targeting the characteristics of \emph{silent bugs} in a widely used framework, TensorFlow through its high-level API Keras. We studied how they manifest in the users' code, what parts of the API are affected and what are their causes. As the main motivation of this work is to study the impact of silent bugs on DL framework users' programs, we validated our study through a questionnaire survey aimed at developers/users of the framework. Fig. \ref{fig:process} illustrates the schematic diagram of our study in this paper.

With more than 111,000 GitHub repositories using TensorFlow, currently, it is one of the most popular ML frameworks \cite{frameworks2021}, hence the impact of elusive bugs could be significant. Keras is an API designed for DL, running on top of TensorFlow. The choice of studying Tensorflow's API Keras is based on several considerations: (1) Tensorflow's Keras API is widely used to implement DL applications due to its simple and rich API, especially for prototyping which is particularly useful in research, (2) being a top-level API using lower level components of TensorFlow, the root cause of the errors should be easier to identify, which simplifies the screening for the root cause if the errors are located in lower-level and (3) there is sufficient information with documented and frequent updates, to ensure identifying silent bugs and their existing fixes.

We hence mined TensorFlow GitHub repository \cite{tensor-repo} for \emph{silent bugs} issues related to its Keras API. We manually analyzed 1,168 issues, gathering and documenting 77 reproducible silent bugs.
Based on the extracted issues, we have identified 7 categories of bugs and 4 levels of bug impact. Moreover, 14 components of the API were identified as responsible for bugs and associated impacts. Finally, we conducted an online survey with 103 DL users of TensorFlow, to assess the relevance and completeness of our identified categories and impact levels.
This paper makes the following contributions:
\begin{itemize}
    \item We conduct a systematic study on \textit{silent bugs} in DL frameworks through manual analysis of 77 bug reports out of 1168 closed issues from the repository of TensorFlow on GitHub,
    \item We provide a classification of bugs based on observed impacts on the users' programs that create, train, or infer DL models,
    \item We identify the components of the Keras API where silent bugs occur as well as root causes if possible using existing taxonomy \cite{jia2021symptoms},
    \item We evaluate the relevance of our findings using a survey of DL users working with TensorFlow,
    \item We make the dataset used in this study publicly available online \cite{rep-package} for other researchers/practitioners to replicate or build upon our work.
    \end{itemize}

\section{Background}

In this section, we briefly introduce TensorFlow/Keras and how their reported bugs are addressed by developers.

\subsection{TensorFlow and Keras}
TensorFlow is an end-to-end, open-source ML platform \cite{abadi2016tensorflow}. Dataflow graphs are used to define the computations, operations, and states of ML algorithms. TensorFlow is an infrastructure layer for flexible differentiable programming that offers gradient computation of arbitrary differentiable expressions and efficient low-level tensor operations on CPU, GPU, or TPU. It is also scalable to various devices. Regarding deployment, it is possible to export graphs to different external runtime environments such as servers, browsers, mobile, and embedded devices.

Keras is a set of DL APIs developed in Python that runs on top of the TensorFlow ML framework \cite{keras}. TensorFlow uses Keras as its high-level API, which provides essential abstractions and building blocks for developing and shipping DL solutions. It was designed and implemented to enable fast experimentation as a key factor in scientific research. Keras is featured as 1) Simple: developing with Keras does not need high skills/knowledge in DL/ML, so the developer can focus on the research problem, 2) Flexible: the progressive disclosure of complexity is adopted in Keras, so it starts with simple workflows that are quick and easy, while arbitrarily advanced workflows are available on the top of the already build simple workflow, and 3) Powerful: industry-strength performance and scalability are offered in Keras, as it has been used by NASA, YouTube, or Waymo. It should be noted that Keras was originally designed to support multiple backends, including TensorFlow, Theano, and PlaidML, however, as of version 2.4, only TensorFlow is supported \cite{Keras-release}.
\subsection{Repairing bugs in TensorFlow/Keras}
The source code of TensorFlow is maintained on GitHub \cite{tensor-repo}. All issues and commits have been reported in this repository respectively since November 2015. Similar to other repositories, users will submit an issue when they face a problem while using the library (e.g., a bug). Such issues contain information for investigating and diagnosing the problem including a brief description of the issue, the OS platform, the buggy TensorFlow version, and the code snippets to reproduce the buggy behavior, as well as a mention of the expected behavior. The developers of TensorFlow (or any other interested users) investigate the issue and potential ways to address it. Not all reported issues are bugs. Usually, the developers first attempt to reproduce the issue and acknowledge the mentioned unexpected behavior. Typically developers and interested users have useful discussions on the issue to understand and then fix the potential bugs. If the issue is identified as a real bug, it will be tagged as a bug and developers start working to address it. Finally, developers will report or mention the link to the fixed version (or similar issues) on the issue asking others to test and acknowledge it. The issue should be closed then. While other issue tracking systems like Jira have more advanced issue trackers labelling issues as ‘‘resolved’’ or ‘‘fixed’’, the issue tracker of GitHub is simpler. Therefore, careful inspection of issues and related discussion is necessary to understand them deeply.


\section{Silent bugs in Keras and TensorFlow}

In this section, first, we describe our methodology to investigate bugs to answer RQ1 and RQ2. The general process consists of three steps: (1) Data collection, (2) Manual analysis, and (3) Labeling and classification as shown in Figure \ref{fig:process}. Then, we discuss our findings.
\subsection{Methodology}\label{sec:meth}
\subsubsection{Data Collection}
We collect issues from the issue tracker of TensorFlow repository over GitHub \cite{tensor-repo}. Various types of issues can be found in such repositories including bug reports, and users' questions (\eg{} misunderstanding documentation). So, it is crucial to identify relevant issues for our study. Since in this study, we aim at investigating the characteristics and consequences of \emph{silent bugs}, we collected closed issues labelled as \enquote{bug}. The reasons for this decision are: 1) Bugs involved in such issues have been acknowledged and then fixed by the framework developers, and 2) Such issues usually include useful information about the bug that helps to understand the bug (\eg{} discussions among users/developers, code/version changes, links to related issues and potential fixes).

We used the GitHub search API \cite{GitSearchAPI} to extract relevant issues. To exclude the search, we have defined the following search criteria accordingly:
\begin{enumerate}
    \item The type of issue must be labelled as “bug” since we are looking for bugs reported by users.
    \item The related component of the issue must be labelled as “keras” since we focus on the Keras API-related bugs.
    \item We extracted only the “closed” issues. This decision aimed to ensure that we would analyze issues that were solved meaning that a fix exists if there the issue relates to a real bug.
\end{enumerate}

So, we filtered issues according to the following query and extracted \textbf{1,168} issues this way:
    \begin{lstlisting}
    is:issue is:closed label:"type:bug" label:"comp:keras"
    \end{lstlisting}

\subsubsection{Manual inspection}
In this paper, our goal is to investigate \emph{silent bugs}, so in our inspection, we first looked for real bugs which are reproducible. Usually, attempts to reproduce the bugs are first realized, \eg{} using a Google Colab a cloud-based Jupyter notebook environment\footnote{\url{https://colab.research.google.com/}}) and providing the link on the issue. We relied on such gists during our inspection as well as acknowledgments that the bug could be reproduced. We then ensure that the bug remains silent during compilation and execution, \ie{} not leading to an error message or crash. Moreover, only fixed bugs were selected during our inspection process since we were looking for closed bug reports. A hint to the fixed version or users' acknowledgment of having the bug fixed was considered as evidence, which generally meant the gist showcasing the bug was executed again and users/developers made sure the bug did not occur with the new version/commit. Each artifact was inspected by reviewing different parts, \ie{} the raised issue, code snippets, discussions mentioned by the owner/other users/developers, suggestions, and recommended fixes. To summarize, we have used the following inclusion criteria to evaluate issues:
    \begin{itemize}
        \item Is it an actual bug? i.e., not a mistake or misunderstanding made by the user. We consider acknowledgement of the framework developers as evidence,
        \item Is the bug reproducible? \eg{} by providing a gist (usually a Google Colab),
        \item Is it \emph{silent}? We regarded as \emph{silent} any bug that did not stop the program, for example not causing the program to crash, hang (with or without an error message), or fail.
        \item Is the bug fixed eventually? The evidence can be mentioning the fixed version, fix commit, or an acknowledgment from the user who raised the issue.
    \end{itemize}
A shared document including the link to all artifacts has been used to make it possible for all authors to work together during the inspection process. It should be noted that each issue is created by a user of the framework who \textit{thinks} a bug was encountered, providing details and a potential gist for reproducing the bug. Then, other users and developers of the framework try to replicate the said bug, and developers would acknowledge whether the issue is indeed a bug or not, simply a user mistake/misunderstanding. We leveraged such a decision to declare that a bug is indeed a bug. Moreover, we made sure of the existence of a fixed version provided by developers, in which the bug would not be observed. As such, the developers' decision (whether it is a bug or not and what is the fix) is our ground truth.

To select relevant bug reports, all issues were initially inspected by three reviewers independently, the first three authors, who have more than 3 years of experience in using TensorFlow. Each reviewer selected issues according to the mentioned inclusion criteria in a separate sheet on our shared document. Afterward, the three reviewers pooled together their results, yielding a score between 0 and 3 for each issue (3 meaning all agreed it satisfies our inclusion criteria and 0 for none of them). \textbf{1,032} bugs received a score of 0, \textbf{41} received a score of 1, \textbf{38} received a score of 2 and \textbf{57} received a score of 3 in this step. Issues with a score of 3 (\ie{} 3 positive votes from reviewers) were directly selected for the next round. Issues with a score of 0 were discarded. All 1-vote bugs were checked in a common meeting by the raters to make sure that there was no overlooked bug, however, none of them were accepted after re-examination. Issues with a score of 2 were further discussed by all reviewers in a meeting, allowing them to resolve conflict and agree on whether or not to select them. \textbf{26} were selected after further discussions and re-inspecting the candidates. Out of \textbf{83} remaining issues, 6 bugs were discarded at the end: 3 turned out to be duplicates (same points but different issue titles), 2 turned out not to be bugs but users’ mistakes, and 1 had got closed after being open for a long time without a consensus on resolving it. In the end, we obtained a total of \textbf{77} bugs. We computed Fleiss’s kappa coefficient \cite{Fleiss71} to assess the inter-rater agreement on whether each bug is silent or not. The coefficient was measured as \textbf{0.705} indicating a good level of agreement \cite{Douglas91}.

All remaining issues were screened by reviewers to collect the following information: version of TensorFlow employed by the user posting the issue (buggy version), link to a gist to reproduce the issue, versions for which the bug is no longer observed (fixed version) and a commit fixing the issue if possible.

\subsubsection{Labeling and Classification}
To answer RQ1, we manually analyzed all issues following an open coding procedure \cite{seaman1999qualitative} similar to previous studies \cite{DL_bugs_2, DL_faults}. This allowed us to find categories based on our observations from the extracted issues. We aimed to inductively create the categories in a bottom-up way by analyzing the bugs.

To analyze problems caused by \emph{silent bugs} and characterize them, we focus on their potential impact on the user's program and the component of the Keras API responsible for the bug. Therefore, we investigated each of \textbf{77} bug reports from two aspects:
    \begin{enumerate}
        \item What is the effect of the bug? That is, what did it cause on the program (\eg{} accuracy loss, wrong UI output, ...).
        \item What are the Keras/TensorFlow components in which silent bugs occur?
    \end{enumerate}
    
To propose a classification label for the bugs' impacts, we used the descriptions of bugs in the issues as well as the gist and/or the example provided to showcase the problem. We call defined labels  \textit{impact scenarios}. Note that, contrary to traditional bugs, \textit{silent bugs} do not exhibit obvious symptoms in the traditional sense, such as crash or hang. As such, what we refer to \enquote{impact} here translates to the effects on the user's program that got noticed through some careful analysis without any clear symptoms. Thus, we will use the term \enquote{impact scenario} to refer to that non-obvious \textit{symptoms} produced by silent bugs. The goal of this RQ is to describe the main consequences of silent bugs in users' programs.

In the case of the library's components for RQ2, we identified modules of the Keras API \cite{kerasAPI} which were involved in the error. Note that some modules affected are hidden in the API the user has access to, but are present in the repository of TensorFlow (\eg{} \enquote{engine} module) \cite{tensor-code}. We considered mainly the API documentation, yet we will also make use of the TensorFlow repository. The buggy component was deduced by going through the bug-fix commit and looking at which files were affected by a change. If no commit reference was available, the reviewer inferred the potential buggy component from the description and discussion in the issue report. We further refined this aspect by labelling the root cause of the bugs based on the fix commits available using the taxonomy proposed by \cite{jia2021symptoms}, to compare the root causes distribution when selecting any kind of bugs or focusing specifically on silent bugs. The goal of this RQ is to highlight components that are more prone to be affected by silent bugs and what are the common root causes of silent bugs.

For both RQ1 and RQ2, we performed the labelling process in three rounds. For the first round, to reduce the subjective bias during the labelling process each reviewer was tasked to independently indicate the effect of bugs (\ie impact scenario) by their own defined descriptive labels. Faulty components were identified independently in this round as well. In the second round, we put individual labelling results together. One of the reviewers checked the results and indicated conflicts. Then two other reviewers went through conflicting issues, commented on the labelling/impact level/faulty component, and proposed a suggestion where applicable. Finally, in the third round, the reviewers went through all artifacts to compare their labelling results, discuss any conflicts, and finalize the categories in a team meeting.

For impact scenarios (categories of bugs), since no classification existed for the problem, we followed an open coding procedure to derive categories in a bottom-up way by analyzing the bugs. We therefore could not report the inter-rater agreement due to the lack of prior-defined categories. However, after resolving conflicts and finalizing labels, we noticed that 31 bugs were classified the same as their final labels by all raters, 33 bugs by two raters, and 13 by one rater respectively, showing an initial agreement of 83\% (64 out of 77).

\begin{figure}
    \centering
    \includegraphics[width=0.7\columnwidth]{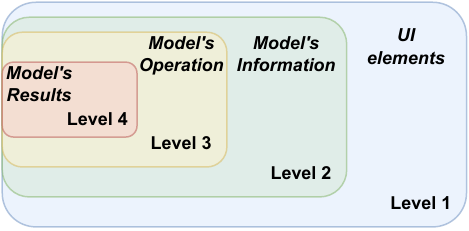}
    \caption{Criteria for evaluating the impact of bugs.}
    \label{fig:vennDiag}
    \vspace{-2em}
\end{figure}

In parallel, during the first round, two reviewers were tasked with analyzing the impact of the bugs in terms of degradation of the ML model's ability to work properly, as well as to analyze potential fixing commits for precise root causes using existing taxonomy \cite{jia2021symptoms}. Given what was observed through the gathered issues, a scale to measure the \emph{threat level} of an issue based on its consequences was designed. This scale was elaborated empirically based on the problems reported by the users and their consequences on the model during discussions between the reviewers in the first round. We then used it as a guideline in the next rounds of labeling and each issue was assigned an impact level. The scale can be found in Figure \ref{fig:vennDiag}. It is divided into four impact levels (1 being the lowest and 4 the highest), each of which encompasses the previous level. For instance, a bug categorized in level $3$ means the bug's effects are predominantly affecting the \textit{Model's Operation}, yet it can also affect both \textit{Model's Information} and \textit{UI Elements}. The defined impact levels are the answers to the question of \enquote{\textit{What is the impact of the bug?}}. They are as follows:
\begin{itemize}
    \item \emph{UI element}: The issue impacts the user interface element, either in a purely cosmetic way or with little to no impact on the information the user receives.
    \item \emph{Model's information}: The issue affects the information the user receives from inside the model, \eg{} shape of a model or reported loss value, but does not modify the way the model works or its results.
    \item \emph{Model's operation}: The issue impacts the model inner mechanisms, \eg{} hyperparameters, layer composition, or model processing mechanism which can lead to some time/speed degradation but only minor to no change on the model outputs.
    \item \emph{Model's result}: The issue impacts the way the model behaves which leads to completely different outcomes compared to known outcomes or the previous version.
\end{itemize}
For example, a bug must impact the information of a model to be considered at least at level 2, higher than bugs that alter UI elements.

For the impact levels, the inter-rater agreement between the two reviewers was calculated as linear weighted Cohen's Kappa \cite{cohen1960coefficient}. The weights are defined as the difference between assigned impact levels where they are different. The rationale is that the threat is increased as the impact level gets higher values. So, we consider that raters assessing an issue of level 1 as level 2 is not as bad as rating an issue of level 1 as level 3. The kappa coefficient obtained is $\textbf{0.758}$ (p-value $<$ 0.05) highlighting a good agreement between raters \cite{Douglas91}. For the root causes labelling, the kappa coefficient is \textbf{0.912} (p-value $<$ 0.05) highlighting an excellent agreement.

All in all, \textbf{77} bugs were classified into \textbf{7} impact scenarios and \textbf{14} components of the API were identified to be affected by the bugs. Each bug was also assigned with one of the impact levels defined in Figure \ref{fig:vennDiag}. We kept all the initial labels and comments on our shared document for any further discussions.

\subsection{Empirical Results}

We provided a replication package of our used material and collected data for reproducibility purposes \cite{rep-package}.

\subsubsection{RQ1: Impact of bugs}\label{sec:taxonomy}

Based on the classification and labelling process described, the identified impact of \emph{silent bugs} is presented as follows. For each category, an example from the replication package \cite{rep-package} illustrating the category is presented. For each of those examples, we give the issue number as well as the verbatim title of the issue along with a description of the problem (\ie what is the expected output and what is the actual output). Those examples allow us to also illustrate the impact level (as defined in the previous section, see Figure \ref{fig:vennDiag}) as well as to detail what consequences could unfold. Each of those samples is the original snippet used to report bugs posted in the GitHub issue tracker of the Tensorflow repository. The categories presented based on their frequency in decreasing order observed in our extracted dataset.

\textbf{1. Wrong calculation (29.9\%):} All bugs modifying the model, as to change the normal way of computation that are not classified in other categories will be classified in this type, like wrong calculation of gradients.

\emph{Issue \#38596 \enquote{Keras fails to account for smaller last batch in loss metric calculation} \cite{issue7}:} In the issue displayed in Figure \ref{fig:wrongcalculation}, the evaluate() function computes the mean loss over all batches in an epoch incorrectly when the dataset size is not evenly divisible by the batch size. This happens for both training and validation loss. Actually, the bug affects the reported epoch loss, but not the training loss used for computing gradient updates. In the provided gist, there are 3 samples in the dataset, and the batch size tested ranges from 1 to 3. The error occurs when the batch size is set to 2. When the batch size is 2, there is one batch of size 2 and one of size 1. If the first batch has a mean loss of 10 and the second batch has a mean loss of 9, then the mean loss over the entire dataset is incorrectly computed as (10 + 9) / 2 = 9.5. However, the correct mean loss over the dataset should be a weighted mean of the batch losses, where the weights are given by each batch size. Thus, the correct mean loss should be (10*2 + 9*1) / (2 + 1) = 9.66, \ie the same as if the batch size is set to 1 or 3. The bug occurred because of an error which made the metric not account for the proper batch size. It was fixed in a subsequent commit. We evaluated this bug having \textbf{Impact: 2} since the model's information is affected but not its functionality. \\

\begin{figure}[H]
\centering
\begin{adjustbox}{left,width=0.9\columnwidth}
\begin{lstlisting}[frame = single]
import tensorflow as tf

X = tf.constant([[1], [2], [3]], dtype=tf.float32)
y = tf.constant([[5], [4], [6]], dtype=tf.float32)
model = tf.keras.Sequential([
    tf.keras.layers.Dense(1, input_dim=1, kernel_initializer='ones', bias_initializer='zeros')])
model.compile(optimizer='sgd', loss='mean_squared_error')
def mse(y, y_pred):
    assert len(y) == len(y_pred)
    return sum((y - y_pred)**2)/len(y)
print('model.evaluate():')
print('- batch_size=1:', model.evaluate(X, y, batch_size=1, verbose=0))
print('- batch_size=2:', model.evaluate(X, y, batch_size=2, verbose=0))
print('- batch_size=3:', model.evaluate(X, y, batch_size=3, verbose=0))
print()
print((mse(X[:-1], y[:-1]) + mse(X[-1], y[-1]))/2)
\end{lstlisting}
\end{adjustbox}

\begin{adjustbox}{left,width=0.9\columnwidth}
\begin{lstlisting}
Output:
model.evaluate():
- batch_size=1: 9.666666984558105
- batch_size=2: 9.5
- batch_size=3: 9.666666984558105

tf.Tensor([9.5], shape=(1,), dtype=float32)
\end{lstlisting}
\end{adjustbox}
\caption{Example of \textbf{Wrong calculation} scenario.}
\vspace{-2em}
\label{fig:wrongcalculation}
\end{figure}

\textbf{2. Wrong parameter setting (20.8\%):} Any bug affecting the expected parameters setting of a function or model's components falls in this category.

\emph{Issue \#31324 \enquote{Passing a Variable as learning\_rate to Adam optimizer does not work as expected} \cite{issue4}:} We refer to this issue as mentioned in Figure \ref{fig:impact3} where passing a variable to learning rate and dynamically changing it will not be registered on the learning rate. As one can see from the provided gist and output, even though the learning rate variable of the optimizer is set to 0, the variable of the trainable layer (\enquote{tf\_a}), , on which the optimizer is applied, is still being modified. This should not be happening since the learning rate is 0 and so the variable value should remain the same since it should amount to modifying \enquote{tf\_a} by 0. We classified this error as \textbf{Impact: 3} as it modified the way of functioning of the network (the learning rate set). In this gist, the learning rate setting is pretty drastic, in practice, the user would probably set something akin to a scheduler and reduce the learning rate slowly. Yet, this issue could lead to potential convergence issues. In practice a \lstinline{LearningRateScheduler} should be defined instead (following good practice), so one could argue the mistake is on the user end, but the fact that no warning is thrown, that no documentation explicitly forbids this and that the mechanism is acceptable by the model shows that is more of a potential issue.\\
\begin{figure}[H]
\centering
\begin{adjustbox}{left,width=0.9\columnwidth}
\begin{lstlisting}[frame = single]
import tensorflow as tf
print(tf.version.GIT_VERSION, tf.version.VERSION)
import sys
print(sys.version_info)
tf_a = tf.Variable(1.0)
print('Variable tf_a initialized to {}.'.format(tf_a.numpy()))
tf_lr = tf.Variable(0.1, trainable=False)
tf_opt = tf.keras.optimizers.Adam(learning_rate=tf_lr)
@tf.function
def train_step():
    with tf.GradientTape() as tf_tape:
        tf_loss = tf_a**2
    tf_gradients = tf_tape.gradient(tf_loss, [tf_a])
    tf_opt.apply_gradients(zip(tf_gradients, [tf_a]))
print('After one step with learning rate {}... '.format(tf_lr.numpy()), end='')
train_step()
print('Variable tf_a is {}.'.format(tf_a.numpy()))
tf_lr.assign(0.0)
for _ in range(10):
    print('After another step, now with learning rate {}... '.format(tf_lr.numpy()), end='')
    train_step()
    print('Variable tf_a is {}.'.format(tf_a.numpy()))
\end{lstlisting}
\end{adjustbox}

\begin{adjustbox}{left,width=0.9\columnwidth}
\begin{lstlisting}
Output:
v2.0.0-beta0-16-g1d91213fe7 2.0.0-beta1
sys.version_info(major=3, minor=5, micro=6, releaselevel='final', serial=0)
Variable tf_a initialized to 1.0.
After one step with learning rate 0.10000000149011612... Variable tf_a is 0.8999971747398376.
After another step, now with learning rate 0.0... Variable tf_a is 0.8004083633422852.
After another step, now with learning rate 0.0... Variable tf_a is 0.7015821933746338.
[...]
After another step, now with learning rate 0.0... Variable tf_a is 0.07624538242816925.
After another step, now with learning rate 0.0... Variable tf_a is 0.005127914249897003.
\end{lstlisting}
\end{adjustbox}
\caption{Example of \textbf{Wrong parameter setting} scenario.}
\vspace{-2em}
\label{fig:impact3}
\end{figure}

\textbf{3. Wrong displayed message (19.5\%):} Any bug that shows information in UI (including console messages) that will deceive the user or affect the user’s understanding of the ML model belongs to this category.

\emph{Issue \#32286 \enquote{Simple model.evaluate() example floods output with = characters} \cite{issue3}:} In Figure \ref{fig:impact1}, an issue is displayed where Keras displays a progress bar that is way too long and which eventually masks part of the UI. The gist used as well as the output UI are shown in the figure. Note that the actual display was drastically shortened as it would not fit in the paper. The progress bar given by Keras is too long during the test phase compared to the training phase, with the UI being flooded with \enquote{=}. A proper amount of \enquote{=} was expected instead, in order to fit the progress bar. The error was linked to a wrong variable being passed to the iteration routine used in the training/evaluation loop of the API. We classified this error as \textbf{Impact: 1} as it just affects the UI of the user. This kind of bug is generally harmless, as they do not really modify anything. However, this type of bug can have a nasty impact when for instance logging results in a file; it would literally fill the log file with thousands of useless characters. If the bug presented only generates a limited number, a case where the number of returned characters would be very high (or infinite) could potentially result in an out-of-memory issue.
\begin{figure}[H]
\centering
\begin{adjustbox}{left,width=0.9\columnwidth}
\begin{lstlisting}[frame = single]
import tensorflow as tf

mnist = tf.keras.datasets.fashion_mnist
(training_images, training_labels), (test_images, test_labels) = mnist.load_data()
training_images = training_images / 255.0
test_images = test_images / 255.0
model = tf.keras.models.Sequential([
    tf.keras.layers.Flatten(),
    tf.keras.layers.Dense(128, activation=tf.nn.relu),
    tf.keras.layers.Dense(10, activation=tf.nn.softmax)
    ])
model.compile(optimizer='adam', loss='sparse_categorical_crossentropy', metrics=['accuracy'])
model.fit(training_images, training_labels, epochs=5)

test_loss = model.evaluate(test_images, test_labels)
\end{lstlisting}
\end{adjustbox}

\begin{adjustbox}{left,width=0.9\columnwidth}
\begin{lstlisting}
Output: Train on 60000 samples
[...]
Epoch 5/5
60000/60000 [==============================] - 2s 38us/sample - loss: 0.2980 - accuracy: 0.8903
10000/1 [================================================
... Literally hundreds of thousands of `=` ...
===========================================] - 0s 26us/sample - loss: 0.2803 - accuracy: 0.8673
\end{lstlisting}
\end{adjustbox}
\caption{A sample of \textbf{Wrong displayed message} scenario.}
\label{fig:impact1}
\vspace{-2em}
\end{figure}

\textbf{4. Wrong save/reload (13\%):} In this category, we classify all bugs that lead to the model, its component, or its functionalities changing either during saving or re-loading. For example, missing a layer after reloading or inconsistent accuracy before and after saving.

\emph{Issue \#42459 \enquote{Accuracy is lost after save/load} \cite{issue1}:} Figure \ref{fig:motivatingExample} shows an issue where reloading saved weights leads to a big drop in accuracy. We classified this error as \textbf{Impact: 4} as it completely changes the results of the model. If the user would not careful, simply using the weights without verification, the model would return mostly incorrect predictions. Moreover, it is important to notice that the weights are not necessarily the first place where one would look to track errors, especially in a deeper model, as no errors/warnings were thrown. This explains the threats of such errors and highlights the importance of multiple checking procedures when using such frameworks.

\textbf{5. Wrong resulting shape (7.79\%):} This category includes any bug making the model output a wrong shape of a tensor without raising an error.

\emph{Issue \#32476 \enquote{Unexpected output shape on custom keras dynamic layer} \cite{issue2}:} The code snippet is shown in Figure \ref{fig:impact2} along with the output. In this issue, Keras returns a bad shape for the user's custom layer. As one can see, the output given by Keras ([(None, (2,))]) differs from the \texttt{compute\_output\_shape} function of the custom Layer ([(None, 2)]) which is the correct expected output. We classified this error as \textbf{Impact: 2} as, while it returns bad information on the model (the shape), it does not actually alter the way the model should work. If the gist is trivial, it is straightforward to envision how this bug could be problematic in a realistic setting; in case a program of the user would lead to poor accuracy (for instance, because of a bad parametrization), the shape would become the main culprit for the bad result, which would possibly take some time until the user finds out that it does not impact its results. The error was reported fixed in the following version but no precise root cause was described.\\
\begin{figure}[H]
\centering
\begin{adjustbox}{left,width=0.87\columnwidth}
\begin{lstlisting}[frame = single]
import tensorflow as tf
import numpy as np

class Example(tf.keras.layers.Layer):
    def __init__(self, **kwargs):
        kwargs["dynamic"] = True
        super(Example, self).__init__(**kwargs)

    def call(self, inputs):
        return inputs

    def compute_output_shape(self, input_shape):
        return [(None, 2)]

inp = tf.keras.layers.Input(batch_shape=(None, 1))
comp = Example()(inp)

model = tf.keras.models.Model(inputs=[inp], outputs=[comp])
model.summary()
\end{lstlisting}
\end{adjustbox}

\begin{adjustbox}{left,width=0.87\columnwidth}
\begin{lstlisting}
Model: "model"
___________________________________________________________
Layer (type)             Output Shape              Param #
===========================================================
input_1 (InputLayer)         [(None, 1)]               0
___________________________________________________________
example (Example)            [(None, (2,))]            0
===========================================================
Total params: 0
Trainable params: 0
Non-trainable params: 0
\end{lstlisting}
\end{adjustbox}
\caption{An example of \textbf{Wrong resulting shape} scenario.}
\vspace{-1em}
\label{fig:impact2}
\end{figure}

\textbf{6. Performance degradation (5.19\%):} Any bug affecting the performance of ML experiments (training or inference) is categorized in this class, \eg{} memory usage or running time (speed). This category does not include changes in the prediction accuracy of the model. This includes the wrong usage of memory but not memory leakage since it is an obvious symptom eventually.

\emph{Issue \#32420 \enquote{TF2.0 - Multiple calls to Keras .fit and .evaluate makes RAM explode and is 25x slower} \cite{issue5}:} In the issue illustrated in Figure \ref{fig:performancDeg}, consecutive calls to either fit() or evaluate() increases the used main memory (RAM) even when calling with the same data. The user noted that such calls take approximately ten times longer than with TF1.x. According to the provided gist, after 3,312 consecutive calls of evaluate() using TF2.0, 1508MB of memory is occupied. Using TF1.x, the memory used is not increased after consecutive calls of evaluate() (staying at 176MB) while the running time was 25 times faster than TF2.0. This information can be accessed by comparing the \textit{timeit} lines as well as the \textit{Increment} columns of both versions in the provided gist. The precise fix was not mentioned but the issue was reported to be fixed in the following nightly build. We evaluated this issue as \textbf{Impact: 3} since the bug significantly affects the operation of the model. Although such consecutive calls are not common in developing DL applications and the bug does not affect the prediction accuracy, the degraded performance may lead to significant problems during deployment, especially in settings where the system's specifications are limited (\eg{} on-board system) or when decision speed is important (\eg{} autonomous driving system).\\

\begin{figure}[H]
\centering
\begin{adjustbox}{left,width=0.9\columnwidth}
\begin{lstlisting}[frame = single]
from memory_profiler import profile
from time import time
import numpy as np
import tensorflow as tf

model = tf.keras.Sequential([tf.keras.layers.Dense(100, activation=tf.nn.softmax)])
model.compile(loss='mse', optimizer='sgd')
@profile
def eval(x, y):
    model.evaluate(x, y)
x = np.random.normal(size=(1,100))
y = np.random.normal(size=(1,100))
for i in range(100000):
    print('iteration', i)
    tic = time()
    eval(x, y)
    print('timeit', time() - tic)
\end{lstlisting}
\end{adjustbox}

\begin{adjustbox}{left,width=0.9\columnwidth}
\begin{lstlisting}
(TF.2.X) Output: iteration 3312
1/1 [==============================] - 0s 4ms/sample - loss: 1.0205
Filename: reproduce_keras_oom.py

Line     Mem usage    Increment   Line Contents
================================================
     9   1508.3 MiB   1508.3 MiB   @profile
    10                             def eval(x, y):
    11   1508.7 MiB      0.4 MiB       model.evaluate(x, y)

timeit 0.09004998207092285

(TF.1.X) iteration 5100
1/1 [==============================] - 0s 1ms/sample - loss: 1.2716
Filename: reproduce_keras_oom.py

Line #    Mem usage    Increment   Line Contents
================================================
     9    176.0 MiB    176.0 MiB   @profile
    10                             def eval(x, y):
    11    176.0 MiB      0.0 MiB       model.evaluate(x, y)

timeit 0.004405021667480469
\end{lstlisting}
\end{adjustbox}
\caption{Example of \textbf{Performance degradation} scenario.}
\label{fig:performancDeg}
\vspace{-2em}
\end{figure}

\textbf{7. Wrong resulting structure (3.9\%):} The category covers all bugs that led to the expected structure of a model being modified, in particular how it is handled by the framework.

\begin{figure}[H]
\centering
\begin{adjustbox}{left,width=0.9\columnwidth}
\begin{lstlisting}[frame = single]
    inner_input = keras.layers.Input((2,))
    dense = keras.layers.Dense(2, activation='relu')(inner_input)
    inner_model = keras.Model(inputs=inner_input, outputs=dense)
    full_input = keras.layers.Input((2,2))
    td_2 = keras.layers.TimeDistributed(inner_model)(full_input)
    model = keras.models.Model(full_input, td_2)
    model.compile('SGD', 'mse')
\end{lstlisting}
\end{adjustbox}

\begin{adjustbox}{left,width=0.9\columnwidth}
\begin{lstlisting}
    model:
\end{lstlisting}
\end{adjustbox}

 \begin{subfigure}{\linewidth}
     \centering
     \includegraphics[width=0.8\linewidth]{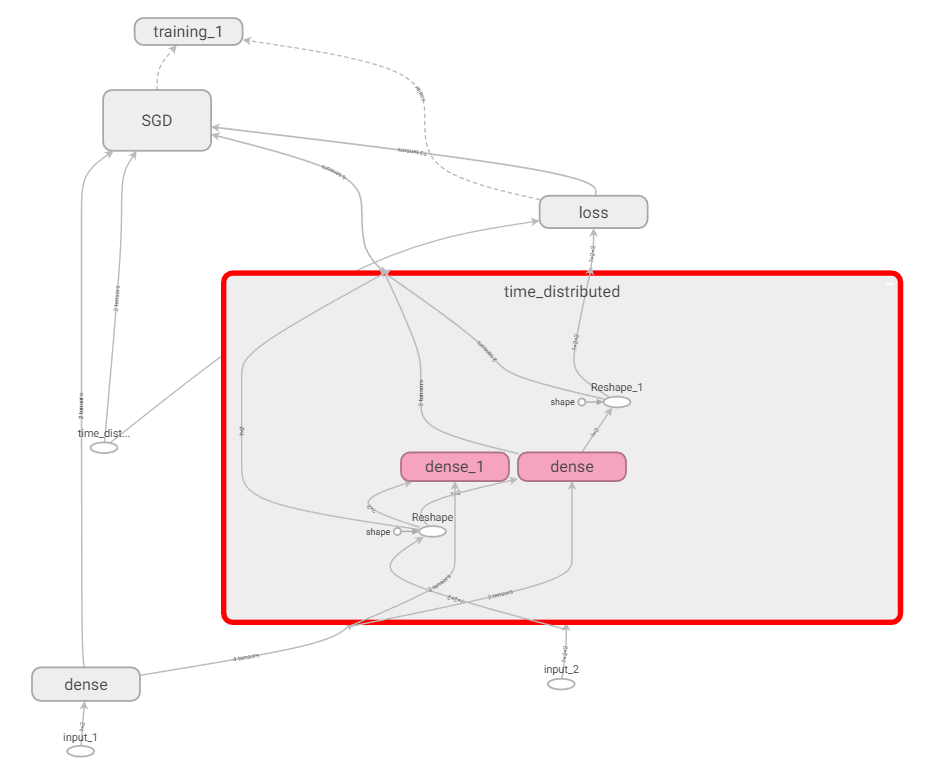}
 \end{subfigure}

\caption{Example of \textbf{Wrong structure} scenario.}
\label{fig:wrongstructure}
\vspace{-3em}
\end{figure}

\emph{Issue \#30486 \enquote{Keras TimeDistributed on a Model creates duplicate layers and is inconsistent with TimeDistributed on a Layer} \cite{issue6}:} As shown in Figure \ref{fig:wrongstructure}, the user attempts to wrap a model in a TimeDistributed layer but this leads to the creation of duplicate nodes in the graph. Following the documentation of the framework, the user ends up with an additional Dense Layer (bottom left in Figure \ref{fig:wrongstructure}) as well as a duplicate Dense Layer in the TimeDistributed layer (\enquote{dense\_1}). The additional layers take redundant memory and are created because the user builds the inner model and then rebuilds it again during building the TimeDistributed. However, the expected behavior for wrapping a model is having a similar graph as it is the case when wrapping a layer, \ie in such a case, there should be only one Dense Layer and it should be within the TimeDistributed layer. This is a typical example of a wrong structure of the model that is silent and the user may not notice it until experiencing a bad accuracy or training issue. The precise fix was not mentioned but the issue was reported to be fixed in the following nightly build. We have evaluated this issue as \textbf{Impact: 2} as it returns false structure information to the user, which can typically lead to confusion or potentially an error. 

\begin{figure}[t]
\centering
\includegraphics[width=0.95\linewidth]{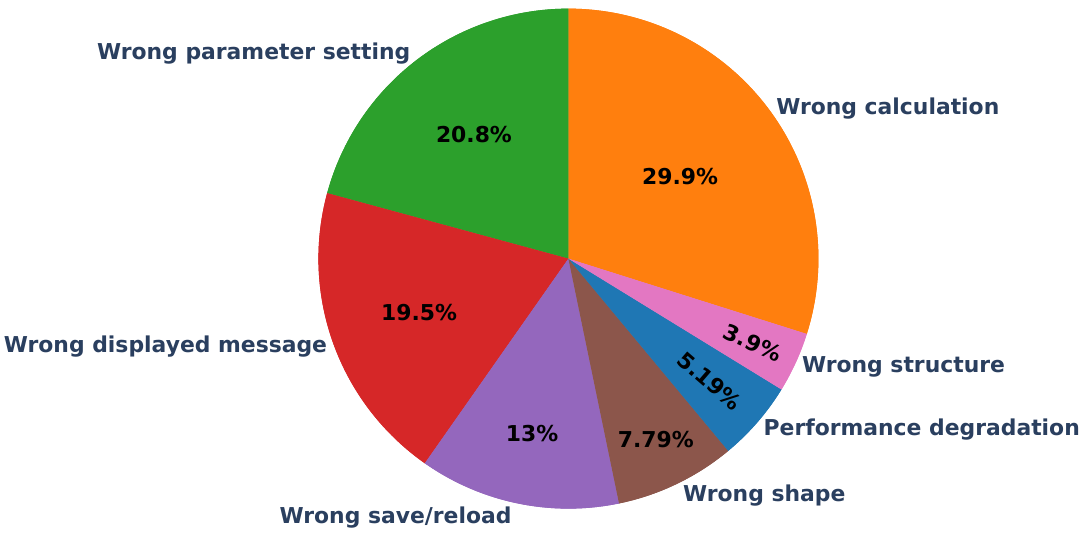}
\caption{Distribution of the impact scenario in the selected issues.}
\label{fig:scenario}
\vspace{-5pt}
\end{figure}

From the extracted data, the distribution of the impact scenarios can be found in Figure \ref{fig:scenario}. \enquote{Wrong calculation} is the most common type with \textbf{$29.9\%$} of all studied bugs. As the internal mechanism and output of DL models mainly rely on their calculation, a flaw in a step of the model's calculation may lead to wrong results in any of the next steps. The two second most widespread types of bugs are \enquote{Wrong parameter setting} (\textbf{$20.8\%$}) and \enquote{Wrong displayed message} (\textbf{$19.5\%$}). On the other hand, \enquote{Wrong structure} has the least frequency with only 3 samples (\textbf{$3.9\%$}).

\begin{figure}[t]
\centering
\includegraphics[width=\linewidth]{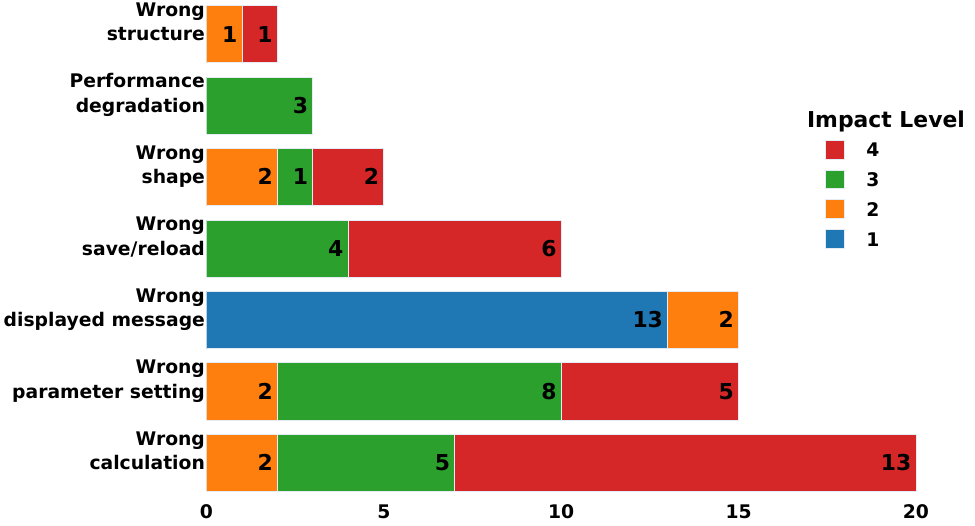}
\caption{Distribution of the impact level based on the scenario affected.}
\label{fig:scenario-impact}
\vspace{-15pt}
\end{figure}

We also show in Figure \ref{fig:scenario-impact}, the cross-check analysis of the impact level given a scenario. The \enquote{Wrong calculation} (\eg{} back-propagation gradients being computed wrongly) has the highest impact with 13 samples of \textbf{Level 4}. The second one is \enquote{Wrong save/reload} (\eg{} weights not being properly set when reloading a model) with 6 high-impact samples. In particular, \enquote{Wrong calculation} and \enquote{Wrong save/reload} bugs represent issues that would drastically affect the results of the model. Moreover, in most cases, they can be only detected through inspection of the model and even that does not guarantee finding the root of the error. As it is expected, \enquote{Wrong displayed message} (\eg{} summary of the model includes missed or incorrect information) is the least influential bug with 13 low-impact bugs.

\begin{tcolorbox}[colback=blue!5,colframe=blue!40!black]
\textbf{Findings 1:} Silent bugs are, in the majority, impactful enough that they modify the expected outcome of the programs which, added to their lack of obvious symptoms, make them particularly dangerous.
\end{tcolorbox}


\subsubsection{RQ2: Faulty components and root causes}

To identify the buggy components of the library we went through the bug-fix commit and looked at which files were affected by a change. If no commit reference was available, the reviewer inferred the potential buggy component from the description and discussion in the issue report. To have a list of Keras API components, we consider mainly the Keras API documentation and the repository of TensorFlow \cite{kerasAPI}. The identified components are:
\begin{itemize}
\item \textbf{Layer:} Layers are the basic building blocks of neural networks in Keras. A layer consists of a tensor-in/tensor-out computation function, held in TensorFlow variables which are the layer's weights.
 
\item \textbf{Regularization:} Regularizers apply penalties on the layer's parameters or its activity during optimization. These penalties are summed into the loss function that the network optimizes.
 
\item \textbf{Callbacks:} They are objects that can perform particular actions at various stages of training (\eg{} at the start/end of an epoch or before/after a single batch). One can use them to log TensorBoard after a batch to monitor training metrics, save the model to disk, early stopping, and gather statistics of a model during training.
 
\item \textbf{Optimizer:} An optimizer is one of the arguments required for compiling a Keras model to optimize the model's parameters during training.
 
\item \textbf{Activations:} Activations can either be used through an "activation" layer or through the activation argument supported by layers. They are applied to the output of tensors to produce the required output.
 
\item \textbf{Metrics:} A metric is used to evaluate the performance of a model. They are similar to loss functions, with the difference that their results are not used when training the model. Any loss function may be used as a metric.

\item \textbf{Model:} These functions are employed to build and use a model including compilation, evaluation, and prediction.
 
\item \textbf{Estimator:} Estimators aggregate the following functions in a single object: training, evaluation, prediction, and export for serving.
 
\item \textbf{Loss:} The purpose of loss functions is to compute the difference between the expected and actual output of a model that should be minimized during training.

\item \textbf{TF:} This encompasses all the functions that are not Keras related but are used by TensorFlow. In general, it is composed of basic operations on the tensors such as \textit{Mean}, \textit{Sum}, or \textit{Slicing}.

\item \textbf{Saving:} This module regroups everything related to saving and loading weights/models to/from the system.

\item \textbf{Engine:} The engine module on which Keras framework is based. It contains most of the basic classes that are used in every other module, \eg{} the \enquote{base layer} class from which all layers are then derived. This module is not meant to be accessible to the user, but as it is the foundation of the whole framework, many issues referenced this module.

\item \textbf{Wrapper:} This encompasses functions to facilitate the usage of other libraries by Keras, mostly Scikit-Learn functions.

\item \textbf{Backend:} The backend module regroups mathematical or utility functions implemented in a way such as Keras can use them more efficiently alongside its other components, as well as a lot of parameter settings for precision calculation.
\end{itemize}

\begin{figure}[t]
\centering
    \includegraphics[width=\linewidth]{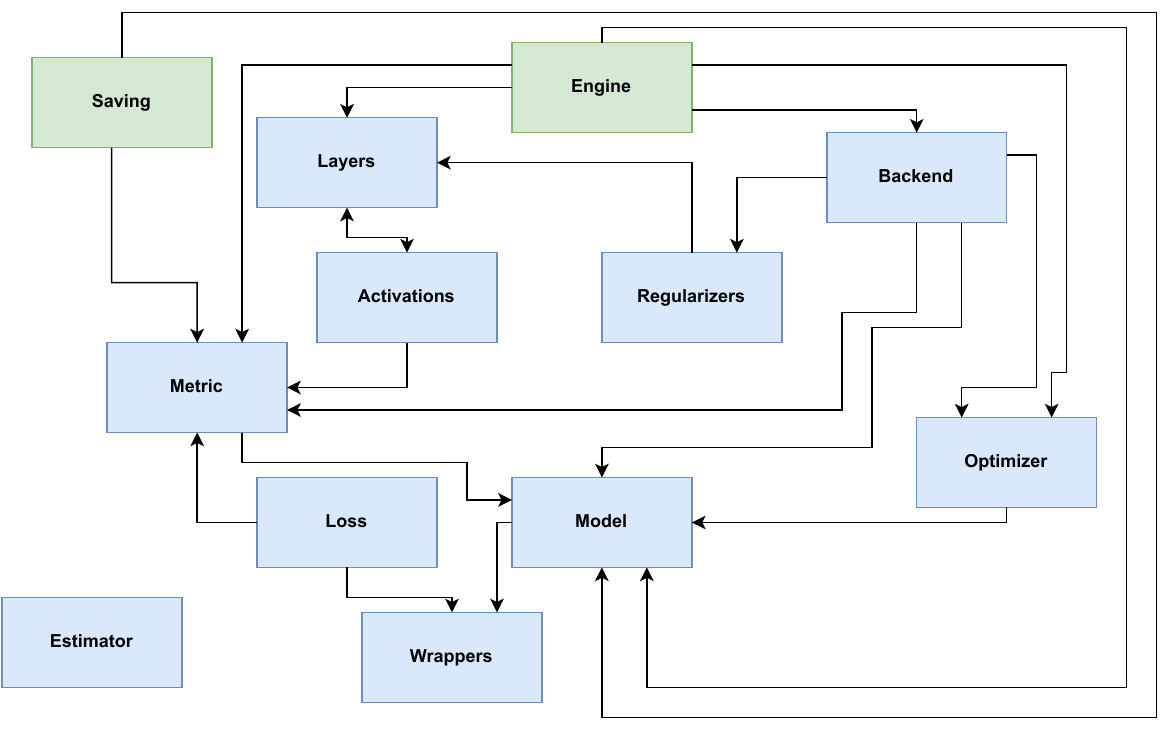}
    \caption{Dependency among the API components. Green components are not supposed to be accessed directly by the user. Connection \enquote{A $\longrightarrow$ B} implies that B is dependent of A.}
    \label{fig:components-relations}
    \vspace{-10pt}
\end{figure}

\begin{figure}[t]
\centering
    \includegraphics[width=1.1\linewidth]{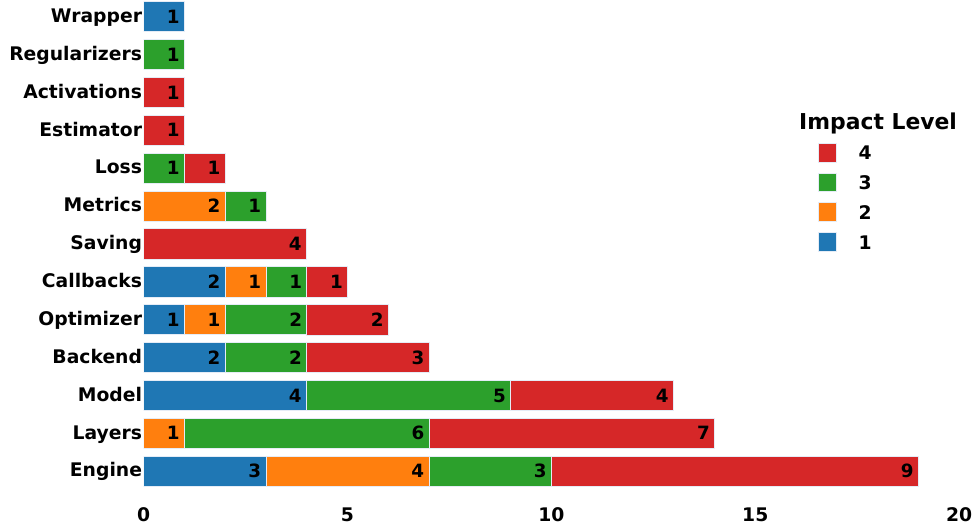}
    \caption{Distribution of the impact level based on the component affected. Bugs affecting the TF component are not displayed here as it is not a part of Keras API.}
    \label{fig:components-impact}
    \vspace{-10pt}
\end{figure}

Figure \ref{fig:components-impact} illustrates the distribution of buggy components across the selected issues along with the cross-check analysis of the impact level given a component affected. The most impacted components of the API are \enquote{Engine} (19 samples), \enquote{Layer} (14 samples) and \enquote{Model} (13 samples). Note that the \enquote{Engine} component is not part of the API documentation as it is a part of Keras backbone and as such should not be accessed directly by the user. Hence, one may expect that this component becomes the most affected one, as it is part of the core of the framework, which all other components use. The two others are among the most used components as they encompass functions and classes that are necessary to define the model layout and inner work. The biggest portion of high-impact scenarios reside in \enquote{Engine} component and \enquote{Wrapper} is the host of only 1 low-impact issue. From a different point of view, all issues in \enquote{Saving} are high-impact ones. Moreover, almost half of the issues in \enquote{Layer} and \enquote{Backend} have high impacts. For completeness sake, Figure \ref{fig:components-relations} shows the relationship between the components we listed. Note that we only display the relationships between the studied components, so we exclude external relations for the sake of readability (that is why for instance \textit{Estimator} has no connections since it does not rely on the listed components but on other TensorFlow parts). 

\begin{tcolorbox}[colback=blue!5,colframe=blue!40!black]
\textbf{Findings 2:} Relatively to the TensorFlow Keras API, silent bugs affect some core components such as the \textit{Engine}, \textit{Layers} and \textit{Models} components. As components of the API are interconnected (see Figure \ref{fig:components-relations}), this in turn can propagate the issue to the other components making it harder to diagnose said silent bugs.
\end{tcolorbox}


For precisely identifying root causes, the two first authors went through the available bug-fix commits and identified root causes independently, using the taxonomy provided by \cite{jia2021symptoms}, before solving conflicts in meetings. If no precise commits were available, the issue was ignored in this part as inferring a precise root cause would not be possible. We ended up with 29 fixing commits to analyze. Our results along with the distribution reported in \cite{jia2021symptoms} are presented in Table \ref{rootcausecompTable}. The reader may refer to \cite{jia2021symptoms} to find full descriptions of the root causes.

While some categories of root causes are similar in proportion (e.g. Type Confusion, Dimension Mismatch and Configuration Error), overall the distribution of root causes we found differs from what was reported by Jia et al. \cite{jia2021symptoms}. Possible explanations can be the fact that we focused on Keras API bugs (that is, bugs happening only in Python language and on the top of TensorFlow) and the fact that we focus more closely on \textit{silent bugs}. Indeed, for those bugs, root causes must produce some changes disrupting normal behaviors \textit{but} without leading to obvious symptoms such as crash or hanging. Thus, for instance, this could, for instance, explain the absence of Referenced types error which is described as \enquote{a bug caused by missing or adding unnecessary include or import statements} or Corner Case \enquote{a bug caused by erroneous handling of corner case}, which would be very likely to lead to some obvious symptoms. Similarly, this can explain the over-representation of Processing, Logic error, and Algorithm categories compared to Jia et al. study, as some bugs in those categories can modify the program's behavior without necessarily leading to a crash, for instance, mask not being applied correctly by the model\footnote{\url{https://github.com/tensorflow/tensorflow/issues/40002}}. Finally, some categories such as Concurrency or Memory might not have appeared because we focused on Keras API (in Python) which might be less likely to produce bugs with such root causes, compared to the Tensorflow part in C++. Indeed, even in Jia et al \cite{jia2021symptoms} study, not only do both categories appear to represent a small percentage of bugs (7 out of 202), but only one of the bugs concerns Python code.

\begin{table}
\caption{Comparison of the root causes distribution in our study and Jia et al. \cite{jia2021symptoms}.}
\label{rootcausecompTable}
\begin{center}
\resizebox{0.6\linewidth}{!}{
\begin{tabular}{|c|c|c|}
\hline

\multirow{3}{*}{\textbf{Root causes}} & \multicolumn{2}{c|}{\textbf{Ratio}}\\
& This study & Jia et al \cite{jia2021symptoms} \\

\hline
\hline
Dimension mismatch & $3.45\%$ & $3.96\%$ \\
\hline
Type confusion & $13.79\%$ & $12.38\%$ \\
\hline
Processing & $\textbf{48.28\%}$ & $\textbf{22.28\%}$ \\
\hline
Algorithm & $10.34\%$ & $2.97\%$ \\
\hline
Logic error & $17.24\%$ & $9.9\%$\\
\hline
Configuration error & $6.90\%$ & $7.43\%$\\
\hline 
Inconsistency & $0.0\%$ & $16.83\%$\\
\hline
Corner case & $0.0\%$ & $15.35\%$\\
\hline
Referenced types error & $0.0\%$ & $4.95\%$\\
\hline
Memory & $0.0\%$ & $2.97\%$\\
\hline
Concurrency & $0.0\%$ & $0.99\%$\\
\hline
\end{tabular}
}
\end{center}
\vspace{-2em}
\end{table}

On top of the root causes, we also tried to label the fix pattern following the categories identified by Jia et al. \cite{jia2021symptoms}. If multiple commits were referenced as fixing the issue, we tried to identify (if possible) which change was linked to fixing the issue and which change was only there to refactor the code to adapt to the fix. If it was not possible or if the fix included multiple fix patterns, we considered the fix pattern as within a ”Multiple” category. The necessity of introducing this category stems from the fact that, while Jia et al. gathered pull requests, we analyzed issue reports (see Section \ref{sec:meth}). As such, while they had access to one fix pattern for one bug, we generally ended up with multiple fix patterns for one bug. For instance, the fix \cite{fix-example} does not fit any of fix patterns mentioned in \cite{jia2021symptoms}: It is not a \textit{Method Replacer} since the method used in the fix does not have the same parameters (the fix calls a more general method) and it is not a \textit{Parameter Modifier} since the added parameter is not part of the same method. After examination, we found that roughly half (53\%) of the issues with precise commits do not have a single fixing pattern: indeed, the commits generally contain multiple modifications that change multiple parts of the code in order to fix the silent bugs. Similarly as root causes, the difference with that study \cite{jia2021symptoms} might be due to the bugs we focus on, silent bugs, that target the TensorFlow’s Keras API, as well as the fact that, in our case, we leveraged the issue report instead of the pull request.

\begin{tcolorbox}[colback=blue!5,colframe=blue!40!black]
\textbf{Findings 3:} Most silent bugs we inspected lack precise fixing commits which makes it difficult to analyze root causes and fixing patterns. Moreover, silent bugs exhibit a different distribution of root causes as well as fixing patterns compared to the general distribution of bugs presented in Jia et al. \cite{jia2021symptoms}. In particular, multiple fix patterns seem to be necessary to fix one silent bug.
\end{tcolorbox}

\section{Validation and relevance assessment}\label{survey}

The goal of this section is to answer RQ3: \emph{To  what  extent  are  the  considered  silent  bugs  relevant for DL users?} To do so, we ran a survey to collect views of DL users about identified categories of silent bugs and the impact scale, assessing their relevance and completeness. In the following, first, the methodology followed to conduct the survey is explained, then the results are presented.
\subsection{Methodology}\label{sec:val_meth}
We devised a survey form using Google Form \cite{googleForm}, a well-known online tool for creating and sharing surveys and quizzes. The survey was organized into five parts and is inspired by similar empirical studies on bugs using surveys as a validation method \cite{nikanjam2021DRL,DL_faults}. First, we asked demographic questions, i.e., current job title and years of experience with TensorFlow. The second part was interested in probing the knowledge of participants about the notion of \enquote{silent bugs}. We first asked them whether or not they were familiar with \enquote{silent bugs}. We then gave them our definition used in this paper and asked them for the perceived relevance. 

In the third part, we wanted to assess the meaningfulness of the scenario categorization we did. As such, for every category, we provide the definition of the category (with an example gist) and a multiple-choice question asking the participant about i) How hard it is to Diagnose, ii) What is their perceived Severity, and iii) How hard it is to fix according to them. The participant is instructed to provide a score on a 5-level Likert scale \cite{oppenheim2000questionnaire}. We then asked them whether or not they were ever confronted with such a type of bug. They could provide feedback if wanted through comments. The main idea of this part is to assess the difficulty of each category to see if we can match the impact classification of Figure \ref{fig:scenario-impact} (that is, if for instance one category is perceived by developers as harder to diagnose or more severe and that we have a prevalence of higher impact level for that category, according to our scale) and see if such a category of bug is rare or not. We then asked developers for any other category they knew and that we did not mention in our classification.

The fourth part aimed to assess the defined impact levels. We gave each level a description along with an example snippet. We then asked them to judge the relevance of the impact level (based on a 5-level Likert scale \cite{oppenheim2000questionnaire}). We finally gave them an impact scale and asked them to evaluate it. The goal of this part was to assess users' opinions: is the impact scale correct about their experience as well as to gauge their comprehension of the scale. 

Finally, with the last question we asked developers whether they thought \enquote{silent bugs} as we presented it to them were more, equally, or less problematic than more traditional bugs. By traditional bugs, we mean the ones returning errors or leading to programs hanging on, \ie not fitting our silent bugs definition. After all the questions were asked, we provided the users with the opportunity to add any optional comment to nuance their views.
We did not ask any questions about the components affected by bugs since it would need requesting participants to study the issue, gist, and bug-fixing commit.

The target group of candidates for this survey is users who use TensorFlow/Keras with some experience in DL programming since we are interested in the impact of bugs on users' code. The first group of candidates came from GitHub. To find participants with a good understanding of DL programming over GitHub, we used its REST APIs \cite{githubREST}. We identified repositories that used TensorFlow (mentioned in their description or Readme). We excluded repositories that were not active since 2019. Finally, we extracted contributors’ emails (mentioned over their profiles) from 12,192 selected repositories. This process left us with 3,582 unique email addresses. For the second group, we referred to issues that we initially collected from the issue tracker of TensorFlow repository over GitHub (1,168 initial bug reports). We extracted emails of users (from their Github profiles) who contributed to the issues, i.e., posting the issues or participating in discussions. We obtained 605 emails in this way. Overall, we successfully distributed the survey participation request to 4187 email addresses. The third group of candidates came from \textit{Reddit} \cite{reddit}. To recruit participants, the questionnaire was posted on two relevant Reddit channels: \textit{deeplearning} and \textit{MachineLearning}. When sending/posting the questionnaire, we explained the purpose, scope, and estimated participation duration (10 minutes) of the survey in a message. Moreover, we asserted that the survey is kept anonymous, but the respondents were able to provide their emails for further communication and to receive a summary of the study. The survey spanned two weeks: the original email was followed by a reminder after one week. The survey form can be found in our replication package along with the anonymous answers that we received \cite{rep-package}.

Finally, we conducted a correlation analysis based on the results of the survey and our own data. we further studied two aspects: 1) whether the impact levels of our bug categories correlate with the perceived impact of the respondents for each category, and 2) whether there was a correlation between the frequency of bugs in our study and the frequency of encounters reported by respondents. To do so, we used Kendall's $\tau$ \cite{Kendall38} which is a rank correlation coefficient measuring the similarity of two rankings of the same data. It compares the average difference between the number of times the ranking is \textit{concordant} (that is, for two data points, both rankings rank the same one data point above the other) and the number of times the ranking is \textit{discordant}.

For the first aspect, we average for each category based on the impact level of each individual bug (so for instance, using Figure \ref{fig:impact1}, \textit{Wrong save/reload} has an average impact level of $3.6$). For the survey, we averaged for each category based on the proportion reported by the respondent in our scale (so for instance, using Table \ref{validationTable}, \textit{Wrong shape} for \textit{Diagnosing} would be $2.42$), and then average the score obtained over \textit{Diagnosing}, \textit{Severity} and \textit{Fixing}.

For the second aspect, we compared the category of bugs we obtained both in our dataset and on the respondents' answers (\textit{Already Encountered} in Table \ref{validationTable}) based on their frequency. Furthermore, we also collected the time to fix each issue in our dataset of bugs in order to add subsequent verification to the second aspect. To do so, we checked the opening date of the issue and considered as the end date the date 1) when the issue was closed or 2) when a fix was acknowledged to be provided by the author of the post and/or a developer. Indeed, the latter often arose as users/developers might have found a fix yet continue to discuss the issue or ask questions before the issue is officially closed. This would have the effect of artificially inflating the time to fix the current bug. For instance, one issue \cite{issue8}, which officially closed in August 2020, was \enquote{considered solved} by the user in June 2020. Once we collected time to fix, we took the median time to fix for each category and compared it to both impact metrics calculated for the first correlation (i.e. our assigned impacts and perceived users' impacts).
\subsection{Validation Results}
The survey was open for two weeks resulting in \textbf{103} answers. Regarding our question on work experience with TensorFlow, $\textbf{9}$ had less than 1 year of experience, $\textbf{51}$ had between 1 and 3 years, $\textbf{36}$ had between 3 and 5 years, and $\textbf{7}$ had more than 5 years. In the following, we discuss the validation results and received comments in detail.

\subsubsection{Notion of Silent Bugs}\label{definitionOfsilent}
When asked, without providing any definitions, whether participants were familiar with the notion of \enquote{silent bugs}, a majority ($\textbf{56.3 \%}$) replied yes. This shows that the term is understood by the community even though it might not be widely used. The unanimity of the respondents ($\textbf{97.1\%}$) found our definition of \enquote{silent bugs} to be meaningful. Hence, the way we defined the term seems to be relevant.

\subsubsection{Categorization of Scenario}

Results of our survey for this part are presented in Table \ref{validationTable}. The first thing we gathered from our survey is, for each scenario, roughly half (or more) of the respondents did not encounter them \emph{to their knowledge}, with \emph{Wrong Structure} being the rarest ($\textbf{28\%}$) and \emph{Performance Degradation} being the most common ($\textbf{55\%}$). Indeed, one main problematic aspect of silent bugs is, that they do not necessarily reveal themselves unless the impact is clearly identifiable or the user is actively looking for them. For instance, one respondent for the \emph{Wrong parameter setting} scenario mentioned that \textit{\enquote{This bug is quite hard to debug without DL knowledge. Mostly, to find them, I had to observe loss changes in different models or different combinations of parameters}}. Indeed, if one just expects this to work, the bug might not be clear during training, especially if training converges to acceptable results as one other respondent pointed out: \textit{\enquote{I may have not noticed it if the model converges anyway}}.
\begin{table*}
\caption{Results of validating survey for the impact scenarios (1: very easy and 5: very hard). Respondents acknowledged that our defined categories, except for \emph{Wrong shape}, \emph{Wrong displayed message}, and \emph{Wrong structure}, are hard/dangerous in terms of diagnosing, severity and fixing.
}
\label{validationTable}
\begin{center}
\resizebox{\textwidth}{!}{
\begin{tabular}{|c|c|c|c|c|c|}
\hline
\multirow{3}{*}{\textbf{Scenarios}} & \multicolumn{4}{c|}{\textbf{Responses}}\\
& Already & Diagnosing & Severity & Fixing \\
& Encountered & 1 | 2 | 3 | 4 | 5 & 1 | 2 | 3 | 4 | 5 & 1 | 2 | 3 | 4 | 5\\

\hline
\hline
\multirow{2}{*}{Wrong Shape} & \multirow{2}{*}{$51\%$} & \multirow{2}{*}{$28\%$ - $30\%$ - $22\%$ - $12\%$ - $8\%$} & \multirow{2}{*}{$8\%$ - $17\%$ - $30\%$ - $28\%$ - $17\%$} & \multirow{2}{*}{$32\%$ - $26\%$ - $24\%$ - $12\%$ - $6\%$} \\
& & & &\\
\hline
Wrong & \multirow{2}{*}{$34\%$} & \multirow{2}{*}{$27\%$ - $19\%$ - $17\%$ - $20\%$ - $16\%$} & \multirow{2}{*}{$23\%$ - $31\%$ - $25\%$ - $14\%$ - $7\%$} & \multirow{2}{*}{$15\%$ - $19\%$ - $27\%$ - $22\%$ - $17\%$} \\
Displayed Message & & & & \\
\hline
Wrong & \multirow{2}{*}{$48\%$} & \multirow{2}{*}{$4\%$ - $17\%$ - $23\%$ - $29\%$ - $27\%$} & \multirow{2}{*}{$4\%$ - $4\%$ - $23\%$ - $33\%$ - $36\%$} & \multirow{2}{*}{$10\%$ - $17\%$ - $31\%$ - $22\%$ - $20\%$} \\
Parameter Setting & & & & \\
\hline
Wrong & \multirow{2}{*}{$47\%$} & \multirow{2}{*}{$14\%$ - $17\%$ - $15\%$ - $25\%$ - $29\%$} & \multirow{2}{*}{$3\%$ - $5\%$ - $17\%$ - $21\%$ - $54\%$} & \multirow{2}{*}{$9\%$ - $12\%$ - $22\%$ - $22\%$ - $35\%$} \\
Save/Reload & & & & \\
\hline
Performance & \multirow{2}{*}{$55\%$} & \multirow{2}{*}{$11\%$ - $13\%$ - $21\%$ - $24\%$ - $31\%$} & \multirow{2}{*}{$5\%$ - $8\%$ - $29\%$ - $33\%$ - $25\%$} & \multirow{2}{*}{$5\%$ - $11\%$ - $20\%$ - $26\%$ - $38\%$} \\
Degradation & & & & \\
\hline
\multirow{2}{*}{Wrong Structure} & \multirow{2}{*}{$28\%$} & \multirow{2}{*}{$7\%$ - $16\%$ - $36\%$ - $25\%$ - $17\%$} & \multirow{2}{*}{$2\%$ - $13\%$ - $32\%$ - $27\%$ - $26\%$} & \multirow{2}{*}{$6\%$ - $20\%$ - $34\%$ - $28\%$ - $12\%$} \\
& & & &\\

\hline
\multirow{2}{*}{Wrong Calculation} & \multirow{2}{*}{$45\%$} & \multirow{2}{*}{$5\%$ - $9\%$ - $32\%$ - $31\%$ - $23\%$} & \multirow{2}{*}{$2\%$ - $13\%$ - $30\%$ - $32\%$ - $23\%$} & \multirow{2}{*}{$5\%$ - $12\%$ - $37\%$ - $25\%$ - $21\%$} \\
& & & &\\
\hline
\end{tabular}}
\end{center}
\label{table:table-results-1}
\vspace{-2pt}
\end{table*}

In general, except for \emph{Wrong shape} and \emph{Wrong displayed message}, there seems to be a consensus among respondents that bugs belonging to our defined categories are rather hard/dangerous (3 or more where 1 indicates the easiest case) in terms of diagnosing, severity and fixing. \emph{Wrong shape} is perceived as easier to diagnose and fix, as one can see from the higher number of respondents who voted for 1 or 2 (both 58\%). However, Severity is also rather high (45\% voted for 4 or 5). On the contrary, for \emph{Wrong displayed message}, diagnosis and severity are rather low with fixing difficulty being neutral. This can be expected, as those bugs are generally easier to notice and/or fix or might not even be fixed in the case of \emph{Wrong displayed message} as they might not be severe enough to be bothered with (\enquote{\textit{Won't fix it myself}}). If we take the average between those three attributes (Diagnosing, Fixing, and Severity) for each category, we find that all categories are above 3.5 (tend to be difficult) with \emph{Wrong save/reload} being the highest, except \emph{Wrong shape}, \emph{Wrong displayed message} and \emph{Wrong structure}. Those observations are consistent with our empirical results presented in Figure \ref{fig:scenario-impact}. Indeed, the categories with the highest number of highly impacting bugs (3 or 4 according to our scale) are the same as those that were deemed the most impactful by respondents when considering the difficulty of diagnosing/fixing and severity. \emph{Performance degradation} was also noted as one highly impacting (i.e., hard to diagnose/fix and severe), which was noted in our figure with all issues of this category being of level 3. The difference we note between received feedback and our results is that \emph{Wrong Shape} might not necessarily be as impactful, yet similarly to \emph{Performance Degradation}, we had too few issues in this category to have a clear conclusion.

Missing categories suggested by the participants were, after analysis, mostly related to one of the categories we introduced or out of the scope of our definition of silent bug (see subsection \ref{definitionOfsilent}). However, we note that some of our proposed categories could be nuanced or defined in more fine-grained terms as they might be too general in their current state, as some developers pointed out. Indeed, bugs affecting control dependency, float approximation, or other \emph{Wrong Calculation} might occult part of the impact of the issue, like one developer commented: \textit{\enquote{when indexing an embedding matrix with negative numbers, TF on GPU does not return an error, but returns vectors made of 0s without notifying the user}}. So, having subcategories could be beneficial
by revealing more information in different subcategories of bugs. One possible category that was brought about but not studied in this paper concerns documentation issues, that is when a behavior is either false or not properly explained in the framework documentation. While such issues are not related to the code, they can still have an impact on the user.

\subsubsection{Categorization of Impact}
Results of our survey for this part are presented in Table \ref{scaleTable}. As one can see a majority of the respondents agree (4 or 5) that all levels are relevant given their definitions. A majority also agree ($\textbf{75.8 \%}$) that our scale is relevant. We note that some respondents pointed out that some levels could be fused together, in particular 3 and 4, for instance, because as pointed out by one comment:\textit{\enquote{Bad learning rates can cause more than just slight alterations in outputs as per my understanding. It can cause instability as well. Hyperparameters like beta1 , lr imo should be level 4.}}. We discriminated between these two levels as it is possible that the affected part of the model structure (even hyperparameter) can have some impact on the model inference, convergence or training but without a high impact on the outcome itself. As such, each bug should be inspected individually to assess the impact level accurately. Overall, we consider following the result of the validation survey that the scale is satisfactory as a first take on quantifying the impact of silent bugs over users' code and finding out which components/impact scenarios are more problematic for the users. We further highlighted this by conducting a correlation analysis between the perceived impacts by respondents and the impact of bugs we collected (see Section \ref{sec:corr}).

\begin{table}
\caption{Results of validation for the impact scale. Respondents have a general agreement on the relevance of all impact levels.
}
\label{scaleTable}
\begin{center}
\resizebox{0.8\linewidth}{!}{
\begin{tabular}{|c|c|c|c|c|c|c|c|}
\hline

\multirow{3}{*}{\textbf{Impact level}} & \multicolumn{5}{c|}{\textbf{Responses}}\\
& Strongly & \multirow{2}{*}{Disagree} & \multirow{2}{*}{Borderline} & \multirow{2}{*}{Agree} & Strongly\\
 & disagree & & & & agree\\

\hline
\hline
Level 1 & $6\%$ & $10\%$ & $18\%$ & $40\%$ & $27\%$ \\
\hline
Level 2 & $7\%$ & $18\%$ & $24\%$ & $36\%$ & $16\%$ \\
\hline
Level 3 & $5\%$ & $14\%$ & $23\%$ & $29\%$ & $29\%$ \\
\hline
Level 4 & $4\%$ & $3\%$ & $11\%$ & $22\%$ & $60\%$ \\
\hline
\end{tabular}
}
\end{center}
\vspace{-2em}
\end{table}

\subsubsection{Comparison with traditional bugs}
We then asked respondents, based on all the questions of the survey and their experience, how problematic they consider silent bugs compared to traditional ones. By traditional bugs, we mean the ones returning errors or leading to programs hanging on. A majority ($\textbf{72.8 \%}$) answered that they found silent bugs to be more problematic. These results highlight the results we described previously of the impacts silent bugs can have on programs of users if not taken into account. 

\subsubsection{Correlation between survey and study conclusions}\label{sec:corr}

With regard to the first aspect of the correlation study: as we have shown, a majority of respondents seem to agree with our impact as well as found silent bugs to be more problematic than traditional ones. They also encountered silent bugs belonging to each category we introduced in our taxonomy (Section \ref{sec:taxonomy}). Following the described methodology at the end of Subsection \ref{sec:val_meth}, for the correlation of the impact levels of bug categories and the perceived impact of the respondents for each category, we found a correlation coefficient of $\tau = 0.51$ (moderate/strong correlation). Thus, the categories with the strongest impact in our evaluation correlate with the perceived impact of the respondents, which further highlights the agreement of the respondents with our scale in the previous section.

With regard to the second aspect of the correlation study: the correlation between the frequency of bugs and the frequency of encounters reported by respondents yielded a correlation coefficient of $\tau = -0.24$ (weak negative correlation). This can be roughly interpreted as the bug categories the most encountered by the respondents tend to be the categories with the least bugs in our dataset. For instance, \textit{Performance Degradation} was the most encountered bugs category for our respondents yet it is among the least represented in our dataset. One possible explanation is that, in general, more impactful bugs type tend to be fixed quicker/be prioritized compared to less impactful bug types.


Thus, we used the median time to fix each category as detailed at the end of Subsection \ref{sec:val_meth} and compared it with the frequency of each category of our dataset as well as respondents' perception. In both cases, the correlation is $\tau = -0.82$ and $\tau = -0.62$ (very strong/strong correlation), which validates our hypothesis (i.e. the more impactful the bugs type the least time to have it fixed). Overall, \textit{Wrong shape} and \textit{Wrong displayed message} are the categories which are the less impactful (both in our assessment and the perceived impacts by users) and which seem to be generally low prioritized when in come to be fixed as they are the categories which take the most time to fix when looking at the median time to fix. On the contrary, \textit{Wrong calculation}, \textit{Wrong save/reload} and \textit{Wrong parameter settings} are the most impactful categories and take the least median time to fix. \textit{Performance Degradation} assessment is different between respondents and our assessment: it's among the bugs categories with the longest fix time and is perceived as the second most impactful category yet it has a medium impact according to our assessment. We summarize that the difference might exist because of the low number of bugs we collected for this category (4) which are however quite different: indeed, one of the four bugs took over a year (the longest of all bugs collected) to be closed compared to the median of 60 days. As such, this category might include very different bugs in terms of complexity to fix and/or priority which could explain why it is the most encountered category for our respondents.

\begin{tcolorbox}[colback=blue!5,colframe=blue!40!black]
\textbf{Findings 4:} Analysis of the respondents' answers validated our silent bugs categorization and impact scale. In particular, respondents largely agreed that \textit{silent bugs} are more problematic than traditional bugs, motivating this paper and further study.
\end{tcolorbox}
\section{Discussion} 

Recent studies by Jia et al. \cite{jia2021symptoms, jia2020empirical} studied symptoms and root causes of bugs inside TensorFlow. While some of their defined categories \textit{could} account implicitly for silent bugs, \eg \textit{Functional error} and \textit{Performance Degradation}, the fact that the issues they use lack users' reports make it hard to diagnose if the bug is silent or not. Moreover, they studied bugs inside the TensorFlow framework in general, without studying specifically the silent bugs category that we have shown can behave very differently compared to traditional bugs. Indeed, \textit{Functional error} is a very broad category, where we showed that it could potentially encompass multiple sub-problems with varying impact over the DL programs: both bugs displaying a wrong message and leading to saving wrongly the weights of a model are functional error, yet the second one is more impactful than the first one. On top of that, while they leveraged pull requests to access bugs, we preferred to use issues reports, as they always come with a proper description of the problem along with a gist provided by the user reporting the bug, which is important in silent bugs cases since they do not exhibit traditional symptoms expected of bugs. Finally, we focused on the API part of Tensorflow and provided an assessment of our results through a survey of DL users using the framework to validate our study. In particular, respondents agreed in the majority that such silent bugs are more problematic than more traditional bugs and so we are convinced they deserve more attention, given their high elusiveness and the harmful impacts we highlighted. 

We showed the issue of silent bugs and the impact they could have on DL programs using Tensorflow Keras API bugs. TensorFlow is one of the most used frameworks by ML practitioners. While users employ the framework \enquote{as it is}, without a proper perspective on what is happening \enquote{under the hood}, we provide evidence that many silent issues can have dramatic impacts, without obvious symptoms, as shown with Findings 1. Although frameworks developers may already be aware of them, it is not the case for all DL framework users. Although there exist different DL frameworks, we surmise silent bugs are likely a non-TensorFlow-specific issue due to the similarities of ML frameworks, which seem to be the case to the light of recent studies \cite{Chen23}.

Finding the actual root causes of silent bugs and how to fix them is not straightforward compared to non-silent ones as silent bugs can affect very different components compared to the impacts on the users' code. As a matter of fact, we are not aware of effective procedures to counter such silent bugs, which are particularly insidious to detect as they do not raise any error, but can lead to highly impacting consequences in DL frameworks. Analyzing the root causes and repair patterns, as well as comparing them following the existing taxonomy \cite{jia2021symptoms} highlighted the difference between general bugs in DL frameworks and silent bugs. Not only was the number of precisely mentioned commits fewer (showing again that finding the actual root causes of silent bugs is harder), but the identified differences between their and our distribution can be explained by differences in the bug collection process (issues vs. pull request, silent bugs vs. general, API vs. general library) which highlights the particular behavior of silent bugs as summarized in Finding 3. In particular, for repair patterns, we found out a majority of said fixes required multiple fix patterns or did not fit any of the repair patterns described by Jia et al \cite{jia2021symptoms}. The more problematic behavior of silent bugs was also confirmed by DL programmers in our survey (Finding 4). We propose the following actionable guidelines for future research on the analysis, detection and localization of silent bugs in the DL framework.

\begin{itemize}
    \item \textbf{Unit test for diagnosing silent bugs:} The fact that multiple silent bugs made it into released versions and that users were confronted with such issues is a testament that, while unit tests are used in DL framework, none of them caught the silent bugs. As such, studying and understanding why and how existing assertion tests could not catch silent bugs is an interesting future work. However for some types of silent bugs, like Wrong save/reload, unit tests could be used simply and effectively: the same structure, parameters, and behavior (like prediction accuracy on a set of benchmark datasets) should be preserved before saving and after reloading the model, which should facilitate designing unit tests. In the general case, we propose using mutations or injected bugs to act as a proxy and evaluate the test suite \cite{Jia22-san, Jia21-icsme}. As the crux of mutation testing is the relevance of said mutations and how they relate to real bugs, leveraging our taxonomy and dataset can help in building mutation operators tailored for silent bugs, which could then be used to assess the relevance of unit tests designed. Another solution would be to use mutations in order to mutate actual existing test cases in order to find bugs, similarly to what was done in \cite{Chen23}.
    \item \textbf{Differential testing for diagnosing silent bugs:} Several techniques have been proposed to deal with bugs in DL libraries/frameworks \cite{Pham19, Wang20, Li23} with some success, though it was shown that such tools generally achieve relatively low test coverage and tend to cover the same part of the code \cite{Chen23}. Most of these techniques use some forms of differential testing over different frameworks using the same models to uncover crashing bugs or inconsistency across frameworks. However, having proper points of comparison can be complicated across frameworks, for instance, it is an open question how to use those techniques to track bugs such as the Wrong Structure (illustrated in Figure \ref{fig:wrongstructure}). Instead, one could leverage different versions of the same framework as an alternative, which reduces problems of inconsistency but increases the risk of the bug not being caught if it is present in all versions used for the test. Another potential solution to this problem is testing each sub-structures individually and then the whole structure at the end. This can be done for the main structures/models in each framework. Taking inspiration from other fields of software engineering where a similar silent bug problem occurs, like in compiler testing with the problem of \textit{miscompilation} (\ie compiler generating a wrong code without error), could foster new techniques adapted to silent bugs in DL frameworks. For instance, Equivalent Modulo Input (EMI) \cite{Le14, Sun16} proved effective at finding miscompilation. The research should be targeted at how to generate equivalent programs, in the EMI sense, in the DL framework case.
    \item \textbf{Identifying root causes:} Even if silent bugs are diagnosed, we have seen that it can be complicated to actually fix them. Finding 3 highlighted that few silent bugs have an actual commit referencing a precise fix. Most of them ended up being fixed in subsequent versions without a precise fix mentioned, probably alongside a non-silent issue. Yet, identifying effectively the root cause could speed up the time to fix the issue as well as provide future insights. One possible solution would be to identify the commit inducing the bug by iteratively testing the version of the framework on the bug revealing code in a similar fashion as \textit{regression testing} \cite{groce2013you}. Automatic bug repair tool \cite{Weimer09, Long16} could be used considering the fact (Finding 3) that fixing said silent bugs seem to require multiple single patches. Moreover, silent bugs also require domain knowledge of the DL framework, which can complexify automatizing the tasks as is the case for numerical bugs (particularly Correctness bugs) \cite{DiFranco17} in traditional Software Engineering.
    \item \textbf{On silent bugs in Tensorflow Keras API:} Finding 2 highlighted that most errors occurred in the \textit{Engine}, \textit{Model} and \textit{Layer} components of the API. As such, we think an extra focus should be directed to testing those components, with \textit{Engine} being especially important as a lot of other modules depend on it. This could have the effect of having some module outputting an incorrect behavior while the real issue is masked by the dependency.
\end{itemize}

\section{Related work} 

\textbf{Silent Bugs in Software Engineering.} Though they might not be defined by the same name, silent bugs were investigated in software engineering. For instance, researchers empirically studied bugs in test code \cite{Vahabzadeh15}. They noted the presence and prevalence of \emph{silent horrors}, that is test code causing a test to miss a bug in the production code. Kouwe et al. studied fault injection to track down \emph{silent failures} and noted through their experiments \cite{Kouwe14} that silent failures are very common. They assessed the impact of silent failures and reported that careful considerations are needed to prevent them from undermining the quality of fault injection results. Moreover, they noted that silent failures were scarcely researched. Similarly, the problem of \textit{miscompilation} in compilers was tackled \cite{Le14, Sun16} with Equivalent Modulo Input, which proved effective at finding such silent defects. However, it is not clear how such an approach could be translated to DL frameworks. Silent bugs in DL frameworks also share some characteristics with numerical bugs (especially the Correctness bugs) \cite{DiFranco17}. The authors note that those types of bugs are extremely challenging to detect, as they require domain knowledge to be fixed, and comment on the utility of Differential Testing to help in detecting them. However, silent bugs in DL frameworks put up additional constraints: contrary to traditional Software Engineering scientific software, DL frameworks are used to implement DL applications on top of them which naturally exhibit some non-deterministic behaviors which are an intrinsic property of DL and which, as a consequence, can make their general testing \cite{Zhang22} as well as diagnosing and fixing of silent bugs harder, for instance by partially obfuscating said bugs. For instance, as we showed with the example in the introduction (Figure \ref{fig:motivatingExample}), it might not be immediately straightforward if it is an error in the saving mechanism, particularly when it happens in a concrete situation (and not in a dummy example as the user provided in the issue report for replication purpose). Accuracy (performance) of the model obtained on the test set is not a reliable indicator \cite{Jia22-san} as it can also vary for many different reasons: some layers (such as Dropout) can have a different effect while training and testing, a different preprocessing of the data can also drastically alter the results. Especially, DL frameworks generally leverage GPU \cite{Nguyen19} which adds some extra complexity through some non-determinism which can affect outputs. As such, it becomes complicated for the users to know whether the error is related to the custom code the users wrote, some feature of DL or indeed a real bug within the DL framework used to develop the DL applications.\\
\textbf{Empirical studies on bugs in DL.} Zhang et al. \cite{DL_bugs_1} published the first empirical study on real-world reproducible bugs occurring in TensorFlow-based software systems including their high-level root causes and symptoms. Then, Islam et al. \cite{DL_bugs_2} extended the investigated cases to include DL software systems written using other competitive DL frameworks such as Pytorch and Caffe, and studied the relationship and the evolution of different types of DL bugs. Last, Humbatova et al. \cite{DL_faults} proposed a taxonomy of real faults that occur in DL software systems. They have not studied the root causes, symptoms, and reproducibility of bugs while they proposed a comprehensive taxonomy of faults in DL programs. However, the focus of all these empirical studies is on bugs in DL programs developed by DL frameworks not the bugs inside DL frameworks. Recently, Jia et al. \cite{jia2021symptoms, jia2020empirical} investigated symptoms, causes, and repair patterns of bugs inside TensorFlow as a typical DL library. While their results are interesting, they have not explored the impact of bugs on the DL programs and the experience of DL users as we do in this paper. Similarly, bugs at different levels (like User-level API and Graph-level implementation) within multiple DL frameworks were studied recently \cite{Chen23}. Bugs in their study that could correspond to silent bugs would belong to the \textit{Incorrect Functionality} and \textit{Poor Performance} categories which make up for 19\% of their analyzed bugs within Tensorflow (i.e. 47 bugs compared to 77 bugs in this study). On the contrary, we delved deeper into the silent bugs as a particular bug category to analyze how they affect the different sub-components of the API, how they are experienced by users (i.e. symptoms) and how they impact the user code. Finally, faults triggered in TensorFlow and similar frameworks have been investigated recently \cite{du2020fault, Du22}. Fault triggers are conditions that activate a fault leading to a failure. The authors aimed to answer research questions about bug type distribution, fixing time, root causes, and regression bugs, by examining bug reports. In particular, they studied bugs from the lens of Bohrbug/Mandelbug and found that most bugs are Bohrbugs with a majority of bugs leading to a crash/hang. On the contrary, the bug category we study in this paper, silent bugs, can be of either category as what distinguishes them is the lack of obvious symptoms.

\section{Threats to Validity}

\textit{Construct validity.} The methodology we used is a potential threat, as the collection and labelling process would have an impact on our study results. We described carefully our approach and extracted the bugs leveraging tags commonly used in Tensorflow's Github repository based on preliminary analysis. Three raters independently analyzed the bugs and discuss the results to avoid biases as much as possible. We only considered closed issues in this study, while open issues might contain some silent bugs, those issues do not provide a solution, commit, or fixed version which could indicate that the bug is: 1) not a user mistake (a real bug), 2) indeed silent, and 3) located in a particular component which is the potential root cause of the bug. As such, to avoid introducing subjectivity and noise in our study, we inspected verified reproducible bugs with a fixed version from closed issues. Regarding categories of scenarios and impact, no previous taxonomy existed and so any classification we would come up with would risk being biased. To mitigate the issue, we adopted an open coding procedure and each rater independently assessed bugs' scenarios/categories with multiple rounds until a consensus was achieved. We further had our categories validated through the DL users survey.

\textit{Internal validity.} The 77 gathered bugs can be a limiting factor. In comparison, other similar studies analyzed 202 bug fixes of Tensorflow \cite{jia2021symptoms}, 175 bugs from Tensorflow applications \cite{Zhang18}, and 250 bugs for TensorFlow \cite{Chen23}, wherein all of them any bugs category in DL frameworks were analyzed. In comparison, we ended up with 77 \textit{silent} bugs as we focus on a particular subcategory of bugs. For example, in \cite{Chen23}, the closest categories to what we identified as silent bugs, i.e. \enquote{Incorrect Functionality} and \enquote{Poor Performance}, only makeup 19\% at most of the 250 bugs for TensorFlow (47), which is less than our study. As we only study one particular subcategory, which is quite hard to diagnose, the number of bugs is comparable. Moreover, with 77 bugs we gathered, we successfully identified and then validated 7 categories of silent bugs.


\textit{External validity.} Because we analyzed only one DL framework, the generalization of the study might be limited. Nonetheless, we believe our study can be generalized to any other DL framework. Similar studies \cite{jia2021symptoms, jia2020empirical} also focused only on TensorFlow as we did. The choice of TensorFlow is motivated by the fact it is currently one of the most used DL frameworks, with the Keras API being a major component of it. Some studies \cite{Chen23} highlighted the commonality in terms of root causes and symptoms of bugs across frameworks, which comforts us in the fact that our results generalize to other frameworks. Nonetheless, even if the phenomenon was not to generalize to other DL frameworks, the impact we established on TensorFlow justifies the importance of the study given the framework’s importance while we believe our categories are general enough and not framework-specific.

\textit{Reliability validity.} To allow other researchers to replicate or build on our research, we provide a detailed replication package \cite{rep-package}.

\section{Conclusion}

In this paper, we conducted the first empirical study to understand the characteristics of silent bugs inside DL frameworks that may result in the design of effective debugging techniques for detecting them. We have showcased \textbf{77} reproducible silent bugs in TensorFlow by inspecting \textbf{1,168} issues from their GitHub repository. The studied bugs were classified into 7 scenarios and 4 impact levels according to the way they can impact DL experiments (designing DL models, training, or inference). We have also identified the framework components that are responsible for the bugs. We then conducted an online survey with \textbf{103} TensorFlow users to validate our results. The participants generally agreed with the significant impact of silent bugs in DL libraries and acknowledged our findings (i.e., categories of silent bugs and the proposed impact scale). Finally, we suggested a set of guidelines to help design effective detection tools for silent bugs. We studied the manifestation of silent bugs in order to help both users and developers to realize the potential danger of such bugs, as well as to provide a first step for developers to identify where they could be located (in the case of Keras/Tensorflow) and how they manifest so they can, hopefully, have an easier time diagnosing them.

Although there exist different DL frameworks, they face similar issues and so we surmise that similar observations could be made. Thus, our results can be used as the basis to study silent bugs in other frameworks. A comparison of the bug types among different frameworks would also be interesting. As another direction for future work, we plan to design detection techniques to help developers identify abnormal behaviors that may be caused by silent bugs and track their root causes.

\begin{acknowledgements}
The authors would like to thank all the participants of the survey who greatly contributed to improving this work with their answers.
\end{acknowledgements}

%
\section*{Data Availability}

All data used for the taxonomy as well as the (anonymized) answers to the user survey are available \cite{rep-package}.

\section*{Conflict of Interest}

 The authors declare that they have no conflict of interest.

\bibliographystyle{spbasic}      
\bibliography{sample-base}   

%
%

\end{document}